\documentclass[iop]{emulateapj}
\usepackage{natbib,xspace,color}
\usepackage{multirow}
\usepackage{verbatim}
\usepackage{rotating}

\newcommand{\lam}{$\lambda$ }
\newcommand{\lala}{$\lambda\lambda$ }
\newcommand{\kms}{\hbox{km s$^{-1}$}}
\newcommand{\ang}{\mbox{\AA} }
\newcommand{\nhI}{$N_{\rm{HI}}$ }

\shorttitle{COS Observations of low-$z$ DLAs and sub-DLAs}
\shortauthors{Battisti et al.}

\begin{document}
\title{The First Observations of Low-Redshift Damped Lyman-$\alpha$ Systems with the Cosmic Origins Spectrograph: Chemical Abundances and Affiliated Galaxies\altaffilmark{1}}
\author{A. J. Battisti\altaffilmark{2}, 
J. D. Meiring\altaffilmark{2},
T. M. Tripp\altaffilmark{2}, 
J. X. Prochaska\altaffilmark{3},
J. K. Werk\altaffilmark{3},
E. B. Jenkins\altaffilmark{4},
N. Lehner\altaffilmark{5},
J. Tumlinson\altaffilmark{6},
C. Thom\altaffilmark{6}
}


\altaffiltext{1}{Based on observations made with the NASA/ESA Hubble Space Telescope, obtained at the Space Telescope Science Institute, which is operated by the Association of Universities for Research in Astronomy, Inc., under NASA contract NAS 5-26555. These observations are associated with program GO11598.}
\altaffiltext{2}{Department of Astronomy, University of Massachusetts, Amherst, MA 01003, USA}
\altaffiltext{3}{University of California Observatories-Lick Observatory, UC Santa Cruz, CA 95064, USA}
\altaffiltext{4}{Princeton University Observatory, Princeton, NJ 08544, USA}
\altaffiltext{5}{Department of Physics, University of Notre Dame, Notre Dame, IN 46556, USA}
\altaffiltext{6}{Space Telescope Science Institute, 3700 San Martin Drive, Baltimore, MD 21218, USA}

\begin{abstract}
We present Cosmic Origins Spectrograph (COS) measurements of metal abundances in eight $0.083<z_{abs}<0.321$ damped Lyman-$\alpha$ (DLA) and sub-damped Ly$\alpha$ absorption systems serendipitously discovered in the COS-Halos survey. We find that these systems show a large range in metallicities, with $-1.10<\rm{[Z/H]}<0.31$, similar to the spread found at higher redshifts. These low-redshift systems on average have subsolar metallicities, but do show a rise in metallicity over cosmic time when compared to higher-redshift systems. We find the average sub-DLA metallicity is higher than the average DLA metallicity at all redshifts. Nitrogen is underabundant with respect to $\alpha$-group elements in all but perhaps one of the absorbers. In some cases, [N/$\alpha$] is significantly below the lowest nitrogen measurements in nearby galaxies. Systems for which depletion patterns can be studied show little, if any, depletion, which is characteristic of Milky Way halo-type gas.  We also identify affiliated galaxies for 3 of the sub-DLAs using spectra obtained from Keck/LRIS. None of these sub-DLAs arise in the stellar disks of luminous galaxies; instead, these absorbers may exist in galaxy halos at impact parameters ranging from 38 to 92 kpc. Multiple galaxies are present near two of the sub-DLAs, and galaxy interactions may play a role in the dispersal of the gas. Many of these low-redshift absorbers exhibit simple kinematics, but one sub-DLA has a complicated mix of at least 13 components spread over 150 km s$^{-1}$. We find three galaxies near this sub-DLA, which also suggests that galaxy interactions roil the gas. This study reinforces the view that DLAs have a variety of origins, and low-redshift studies are crucial for understanding absorber-galaxy connections.
\end{abstract}

\section{Introduction}
Quasi-stellar object (Quasar or QSO) absorption lines have long been used to probe the intergalactic medium and also the interstellar medium (ISM) of galaxies at all redshifts. The damped Lyman-$\alpha$ systems (DLAs) (log \nhI $\ge 20.3$) are thought to contain most of the neutral gas in the Universe at high-redshifts \citep{Pro09,Not09}, with sub-DLAs ($19.0\le \log$ \nhI $< 20.3$) also believed to have significant fraction at high-redshift \citep{Per05}. At lower redshift, a survey for sub-DLAs at $z\sim 2$ by \citet{OMea07} found that they contain $\sim$15\% of the Universe's H atoms. The large H~I column densities of both DLAs and sub-DLAs are typical of what is seen in the disks and nearby environments of galaxies. As such, these systems are believed to be strongly related to galaxies at their observed redshift. At $z\sim 0$, DLAs have been associated to galactic ISM \citep{Zwa05}. These systems provide a unique means to probe galaxies over $>90\%$ of the cosmic history, without the biases introduced in magnitude-limited samples of galaxies studied in emission. 

Since these systems have such large column densities, the majority of the H is expected to remain neutral due to self-shielding. These high column densities also make absorption lines from metal atoms more detectable. Any metals inside one of these clouds will be dominantly in states with ionization potentials greater than that of H$^0$, since photoionization effects will be negligible. For sub-DLAs, the lower column densities make these photoionization effects more significant and corrections are typically applied through use of ionization models. 

DLAs are of particular importance to chemical evolution studies. Measurements of element abundances in such systems provide insights into the production of metals through cosmic time. This task is complicated by the large scatter in metallicities seen at any given redshift, due to the random sampling of galaxies with different masses and different environments. However, this is actually beneficial, since the sample will naturally probe a more representative range of all metallicities and provide an unbiased description of chemical enrichment. Despite the large scatter, using the large sample size of DLAs measured at high-redshift, \citet{Pro03b} found evidence for an increase in elemental abundances with increasing cosmic time. This picture is consistent with our expectation that the Universe becomes more metal-rich as it evolves. Interestingly, many DLAs and sub-DLAs appear to be underabundant in N with respect to the other metals in these systems \citep{Jen05,Tri05,Pet08}, even at low-redshifts. Observations of N in DLAs and sub-DLAs are important for the study of N production and galactic chemical evolution in general \citep{Cen03,Hen07,Pet08}. 

Unfortunately, relatively few measurements of low-redshift DLAs and sub-DLAs have been made \citep{Kul05,Rao05} because the atmospheric cutoff of light below $\sim$3000\ang makes ground based observing of the Lyman-$\alpha$ line impossible for redshifts $z<1.65$. With the Cosmic Origins Spectrograph (COS) aboard the Hubble Space Telescope (HST), we are able to study these low-redshift systems with unprecedented efficiency. This instrument provides us with a great opportunity to expand this small sample. These systems are crucial for a clear understanding of chemical evolution since $z<0.5$ spans $\sim$40\% of the age of the Universe.

Although much can be learned from absorption line data, the difficulty in the interpretation of these systems stems from the lack of information about the host galaxies of DLAs and sub-DLAs at all redshifts. The number of DLAs with spectroscopically confirmed hosts is very small and this is in part due to the much lower number of low-redshift DLAs compared to high-redshift. At high-redshift, the cosmological dimming of the surface brightness of galaxies ($\mu\propto(1+z)^{-4}$) makes galaxy detections more difficult. Low-redshift DLAs present a unique opportunity to study the absorber-galaxy connection, since the host will be more easily detected.

In this paper, we report on 3 DLAs and 5 sub-DLAs over a redshift range of $0.0830<z_{abs}<0.3211$. In \S2 we give background information on the discovery of these systems and on the observations. In \S3 we describe our techniques for determining column densities from absorption lines. In \S4.1 we report the total column densities and abundances of each system, correction for photoionizing effects in the sub-DLA cases. In \S4.2 we report the depletion patterns of systems with a sufficient number of detected lines for analysis. In \S4.3 the metallicities of our absorbers will be compared to those in the literature and also with chemical evolution models. In \S5 we discuss nitrogen measurements for each system and possible implications on nitrogen nucleosynthesis. In \S6 we identify affiliated galaxies for 3 of the sub-DLAs using observations from a redshift survey with Keck/LRIS. Conclusions are summarized in \S7. Throughout this paper, we adopt a cosmological model with $\Omega_m=0.30$, $\Omega_{\Lambda}=0.70$, and $H_0=70$ \kms\ Mpc$^{-1}$.

\section{Program summary and Observations}
The DLAs and sub-DLAs in this sample were discovered serendipitously as part of programs GO11598 and GO12248 (PI: Tumlinson), programs focused on studying multiphase baryons in halos of $L$ $\geq$ $L^{\star}$ galaxies and dwarf galaxy halos, respectively. Early results from these programs can be found in \citet{Tho11}, \citet{Tum11a} and \citet{Tum11b}. The names and basic properties of these absorbers are presented in Table \ref{Tab:Prop}. The DLA with the highest column density in our sample, SDSS J1009+0713, was presented in detail in \citet{Mei11} along with basic information of the other systems. The details of our COS observations are summarized in Table \ref{Tab:observ}. Follow-up Keck/HIRES \citep{Vog94} observations were acquired for three of our absorption systems to see more detailed component structure, since COS has lower resolution (R$\sim$20,000) compared to HIRES (R$\sim$44,000), and those are also summarized in table \ref{Tab:observ}. A more detailed description of the COS observations for this project along with its data reduction methods were outlined in \citet{Mei11}.

\begin{table}
\caption{Properties of the DLAs and sub-DLAs \label{Tab:Prop}}
\begin{tabular}{cccc}
\hline\hline 
QSO & $z_{ems}$ & $z_{abs}$ & log \nhI \\
 & & & (cm$^{-2}$)\\ \hline
SDSS J092554.70+400414.1 & 0.471 & 0.2477 & 19.55$\pm$0.15 \\
SDSS J092837.98+602521.0 & 0.295 & 0.1538 & 19.35$\pm$0.15 \\
SDSS J100102.55+594414.3 & 0.746 & 0.3035 & 19.32$\pm$0.10 \\
SDSS J100902.06+071343.8 & 0.456 & 0.1140 & 20.68$\pm$0.10 \\
SDSS J143511.53+360437.2 & 0.429 & 0.2026 & 19.80$\pm$0.10 \\
SDSS J155304.92+354828.6 & 0.722 & 0.0830 & 19.55$\pm$0.15 \\
SDSS J161649.42+415416.3 & 0.440 & 0.3211 & 20.60$\pm$0.20 \\
SDSS J161916.54+334238.4 & 0.471 & 0.0963 & 20.55$\pm$0.10 \\ \hline
\end{tabular}
\textbf{Note.} Excluding J1001+5944, measurements taken from Meiring et al. (2011). Second and third columns denote the redshift of QSO emission and the redshift of the absorption system, respectively.
\end{table}

\begin{table}
\caption{Summary of COS and Keck observations \label{Tab:observ}}
\begin{tabular}{ccccc}
\hline\hline 
QSO & Obs. Date & Grating & Central \lam & Exp. Time\\ 
 & & & (\ang) & (s) \\ \hline
J0925+4004 & 2010-06-15 & G130M & 1318, 1327 & 2176, 1589 \\
 & 2010-06-15 & G160M & 1600, 1623 & 1248, 3055 \\
J0928+6025 & 2010-03-06 & G130M & 1291, 1309 & 1163, 1148 \\
 & 2010-03-06 & G160M & 1600, 1623 & 1547, 1505 \\
J1001+5944 & 2011-03-18 & G130M & 1291, 1309 & 1618, 1682 \\
 & 2011-03-19 & G160M & 1600, 1623 & 2710, 2491 \\
J1009+0713 & 2010-03-29 & G130M & 1291, 1309 & 2191, 1497 \\
 & 2010-03-30 & G160M & 1600, 1623 & 2007, 2002 \\
J1435+3604 & 2010-05-22 & G130M & 1291, 1327 & 1788, 1678 \\
 & 2010-05-22 & G160M & 1577, 1623 & 2212, 2212 \\
J1553+3548 & 2010-10-01 & G130M & 1291, 1309 & 1600, 1623 \\
 & 2010-10-02 & G160M & 1600, 1623 & 1436, 1401 \\
J1616+4154 & 2010-07-19 & G130M & 1291 & 3644 \\
 & 2010-07-19 & G160M & 1600, 1623 & 2437, 2167 \\
J1619+3342 & 2010-08-19 & G130M & 1291, 1309 & 2358, 2989 \\
 & 2010-08-19 & G160M & 1600, 1623 & 4394, 4395 \\ \hline
J0928+6025 & 2010-03-26 & HIRES & - & 3000 \\
J1009+0713 & 2010-03-26 & HIRES & - & 3600 \\
J1619+3342 & 2010-09-02 & HIRES & - & 3600 \\ \hline
\end{tabular}
\end{table}

After the data reduction, the spectra were binned by 3 pixels, as the raw COS data are oversampled with a $\sim$6 pixel wide resolution element. All subsequent measurements and analysis were performed on the binned spectra. The binned spectra have a resolution of $\sim$15 \kms\ per resolution element. The signal-to-noise ratio (S/N) of these COS spectra range from 7 to 15 per resolution element. 

\begin{figure}
\plotone{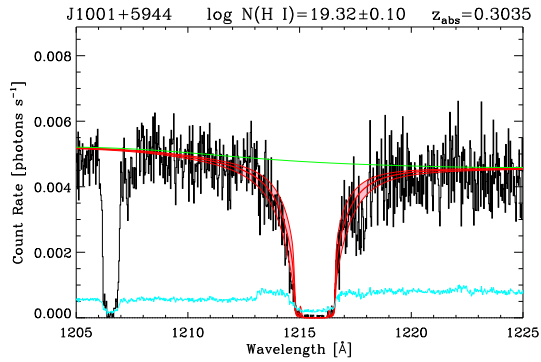}
\caption{Fit to the Ly-$\alpha$ line for J1001+5944. The solid green line represents the adopted continuum and the red lines are the best-fit model with column densities 1$\sigma$ above and below the best-fit value. The 1$\sigma$ flux uncertainty is shown in cyan below the data. \label{Fig:Lyalpha}}
\end{figure}

Accurate column densities for the H~I \lam 1215.67 absorption line in DLAs and sub-DLAs are measured by fitting the damping wings in the profile. Fits to the Ly-$\alpha$ lines all of our systems were presented in \citet{Mei11}, with the exception of J1001+5944 which is presented in Figure \ref{Fig:Lyalpha}. The Ly-$\alpha$ lines for all systems, except J1001+5944, were fitted using a single component with the continuum and profile fit simultaneously. The redshifts of the systems were determined from the metal lines, which with narrow features provide for a more accurate redshift determination. For J1001+5944, we used the four components found in O~I lines to fit the Ly-$\alpha$ line. In total, we find 3 DLA systems and 5 sub-DLA systems in these COS data. Absorption redshifts and neutral hydrogen column densities for the systems were given in Table \ref{Tab:Prop}. Errors on \nhI measurements given in Table \ref{Tab:Prop} have been estimated by eye and are dominated by the continuum placement uncertainty.

\section{Analysis}
Here we present a detailed description of our methods for determining the column densities from observed absorption features. The column densities for unsaturated lines were estimated from a multicomponent Voigt profile fit using the fitting code of \citet{Fitz94}. This code minimizes the $\chi^2$ value between the data and the synthetic Voigt profiles convolved with the COS line spread functions (LSF)\footnote{Tabulated LSFs for COS are available at \\ \texttt{http://www.stsci.edu/hst/cos/performance/spectral\_resolution/}} at the nearest tabulated wavelength \citep{Gha09}. The column density errors for our profile fits are formal 1$\sigma$ uncertainties from the $\chi^2$ minimization and do not include contributions from the uncertainties in the continuum placement. The fitting procedure for each system was to select the species with the most extensive number of observed lines, preferably over a large range of oscillator strengths, $f$, to reduce the possibility of saturation effects. Then we use the velocity centroids and Doppler parameters, $b$, from the best fit of this species for the fits to other lines and only allow the column densities in each component to vary (see appendix for more details). We fit multiple transitions simultaneously whenever possible. Lines showing evidence of being strongly blending with another line were not used for the profile fit of that transition. The COS spectrum was also allowed to shift in velocity space to account for calibration issues in the instrument, which can cause offsets between various transitions of order $\pm$10\kms\ \citep{Sav11}. If Keck data was available, then we determined the component structure by fitting the Mg~I \lam 2852.96 line. The Keck/HIRES spectra were convolved with a Gaussian LSF with FWHM$=2.5$ pixels. 

It is worth noting that the limited resolution of COS will prevent it from detecting components that Keck/HIRES can resolve. This will not dramatically affect column density measurements but does hinder our understanding of the kinematics of these systems. Component fits using only the COS spectrum should be treated with caution for analysis of kinematic structure as they likely do not reflect the true number of components. 

Our profile fitting approach is chosen over apparent optical depth (AOD) methods whenever possible because the AOD is more prone to systematic errors from mild saturation and noise; both of which are concerns in these COS spectra due to the relatively low S/N and broad LSF of the COS spectrograph. By fitting multiple transitions, over a range of $f$ values, we have a better chance to overcome these effects. We assume that species with similar ionization potentials can be fit using identical $b$ values because the resolution of COS is not high enough for small changes in $b$, caused by the different atomic masses, to be significant. One also expects different species with similar ionization potentials to exist in similar regions of a cloud and thus have similar velocity components. Intermediate-ions, such as Fe$^{++}$ and Si$^{++}$, appear to have velocity structure that is indistinguishable from the low-ion profiles in most systems and so it has been argued that the two regions are physically connected or are partially mixed \citep{Des03}. For this reason we also choose to fit the intermediate-ions using the same parameters from the fit to the lower ions. Rest frame wavelengths and $f$ values for all lines were taken from \citet{Mor03}. 

Lines were determined to be saturated in two ways: (1) We compared the total column density found from profile fitting with the apparent column density found from direct integration of the line using the AOD method of \citet{SS91}. Cases for which apparent column density measurements and profile fitting measurements were inconsistent within their uncertainties were determined to be saturated. (2) For species with several lines detected over a range of $f$ values, lines with larger $f\lambda$ values will give lower values of apparent column density than lines with smaller $f\lambda$ values if saturation occurs. When a line is determined to be saturated we report a lower limit based on the apparent column density found for the line of that species with the lowest $f\lambda$ value (i.e. the least saturated line). 

Since the component structure in the more highly ionized gas traced by the CIV \lala 1548.20, 1550.77, Si~IV \lala 1393.75, 1402.77 and O VI \lala 1031.92,  1037.61 lines are often different than that of the low ions, we have measured the column density of these species by direct integration of the line using the apparent column density.

We present the adopted velocity components and $b$ values for each system as well as the column densities of detected species at each of these component in the appendix. The corresponding fits using these parameters are shown in Figure \ref{Fig:J0925vel}-\ref{Fig:J1619vel}.

\section{Chemical Abundances}
\subsection{Total Column Densities and Abundances}

\begin{table*}
\caption{Total logarithmic column densities, log $N_{\rm{X}}$ (cm$^{-2}$), for each absorber \label{Tab:TotCol}}
\begin{tabular}{cccccccccc}
\hline\hline 
X & J0925+4004 & J0928+6025$^a$ & J1001+5944 & J1009+0713$^b$ & J1435+3604 & J1553+3548 & J1616+4154 &  J1619+3342 \\ \hline
HI  & 19.55$\pm$0.15 & 19.35$\pm$0.15 & 19.32$\pm$0.10 & 20.68$\pm$0.10 & 19.80$\pm$0.10 & 19.55$\pm$0.15 & 20.60$\pm$0.20 & 20.55$\pm$0.10 \\
CII & $>$15.18       & $>$14.91       & $>$15.06       & $>$15.90       & $>$14.52       & $>$14.35       & $>$15.15       & $>$14.38 \\
CIII & $>$14.17      &  \nodata       & $>$14.52       &  \nodata       & $>$14.33       &  \nodata       &  \nodata       &  \nodata \\
CIV$^c$ &  \nodata   & $>$14.1        &  \nodata       & 13.86$\pm$0.20 &  \nodata       & 13.99$\pm$0.08 &  \nodata       & 13.28$\pm$0.08 \\
CII$^*$ & $<$14.18   & $<$13.82       & $<$13.73       & $<$13.70       & $<$13.68       & $<$13.46       & 13.95$\pm$0.08 & $<$13.12 \\ 
NI  & 14.75$\pm$0.04 & 14.10$\pm$0.09 & 13.65$\pm$0.17 & 15.11$\pm$0.10 & 14.60$\pm$0.14 & $<$13.74       & $>$15.35       & 14.64$\pm$0.12 \\
NII & $>$14.95       & $>$14.85       & $>$14.95       &  \nodata       &  \nodata       & 14.16$\pm$0.07 & $>$15.17       & $>$13.92 \\
OI  & 15.95$\pm$0.09 & $>$15.08       & 15.64$\pm$0.02 & $>$16.00       & $>$15.58       & $>$14.56       & $>$15.45       & $>$14.80 \\
OVI$^c$ &  \nodata   &  \nodata       & 14.34$\pm$0.05 &  \nodata       &  \nodata       &  \nodata       &  \nodata       &  \nodata \\
MgI  &  \nodata      & 12.70$\pm$0.06 &  \nodata       & $>$12.85       &  \nodata       &  \nodata       &  \nodata       & 12.40$\pm$0.14 \\
MgII &  \nodata      & $>$13.99       &  \nodata       & $>$15.35,$<$16.0 &  \nodata     &  \nodata       &  \nodata       &  \nodata \\   
SiII & 14.62$\pm$0.06 & $>$14.39      & 14.73$\pm$0.03 & $>$15.00       & $>$14.11       & 14.22$\pm$0.05 & $>$15.08       & $>$13.93 \\  
SiIII & $>$13.74     & $>$13.77       & $>$14.00       &  \nodata       & $>$13.41       & $>$13.3        & $>$13.93       & $>$13.24 \\
SiIV$^c$ & 13.54$\pm$0.11 & 13.86$\pm$0.12 &  \nodata  &  \nodata       & 13.20$\pm$0.13 & 13.59$\pm$0.08 &  \nodata       & 13.06$\pm$0.05 \\
PII & $<$13.50       & $<$14.17       & 12.81$\pm$0.07 & $<$13.80       & $<$12.79       &  \nodata       & 13.46$\pm$0.10 & 13.16$\pm$0.19 \\
SII & $<$14.72       & $<$14.65       & $<$14.53       & 15.25$\pm$0.12 & 14.60$\pm$0.12 & $<$14.24       & 15.37$\pm$0.11 & 15.08$\pm$0.09 \\
CaII &  \nodata      & 12.81$\pm$0.06 &  \nodata       & 11.96$\pm$0.08 &  \nodata       &  \nodata       &  \nodata       & 12.42$\pm$0.02 \\ 
TiII &  \nodata      & $<$11.94       &  \nodata       & 12.70$\pm$0.03 &  \nodata       &  \nodata       &  \nodata       & 11.90$\pm$0.04 \\
FeII& 14.22$\pm$0.09 & 14.90$\pm$0.08 & 14.30$\pm$0.04 & 15.29$\pm$0.17 & 14.20$\pm$0.08 & 14.01$\pm$0.07 & 15.02$\pm$0.05 & 14.38$\pm$0.15 \\ 
FeIII & $<$14.23     & 14.59$\pm$0.11 & 14.14$\pm$0.09 &  \nodata       & $<$13.69       &  \nodata       & 14.51$\pm$0.04 & 13.95$\pm$0.21 \\ 
NiII & $<$14.41      & $<$13.67       & $<$14.23       & 13.93$\pm$0.18 & $<$13.99       & $<$13.92       &  \nodata       & $<$13.53 \\\hline
\end{tabular}
\newline$^a$ The COS lines were determined using the apparent column density. \newline $^b$ Measurements taken from Meiring et al. (2011). \newline $^c$ Higher ions measured using apparent column density.
\end{table*}

Total column densities measurements were determined by summing the individual components from Table \ref{Tab:Param} or from apparent column density measurements and are listed in Table \ref{Tab:TotCol}. Upper limits to undetected lines were derived from a 3$\sigma$ upper limit on the equivalent width, assuming the line is on the linear part of the curve of growth.

The absolute abundances were measured with respect to solar using the following formula:
\begin{equation}
\rm{[X/H]=log[N(X)/N(H)]}_{\rm{DLA}}-\rm{[X/H]}_{\sun}+\epsilon_{\rm{X}}
\end{equation}
where $\epsilon_{\rm{X}}$ is an ionization correction factor applied to the sub-DLAs. Solar system reference abundances ([X/H]$_{\sun}$) were taken from \citet{Asp09} adopting the photospheric values for C, N and O and meteoric values for all other elements. In these systems the dominant species is the neutral state for O and N and the first ions for C, Mg, Si, P, S, Ti, Fe, and Ni. Abundance measurements are not made for Ca, since the dominant ion in this case is Ca$^{++}$, which doesn't have lines in our observing window. For saturated and unobserved lines we present lower and upper limits to abundances, respectively, using the AOD method. In Table \ref{Tab:Abund} we show the resulting abundances for each absorber without applying $\epsilon_{\rm{X}}$.

\begin{table*}
\caption{Abundances, [X/H], for each absorber with respect to solar \label{Tab:Abund}}
\begin{tabular}{cccccccccc}
\hline\hline 
X & J0925+4004$^a$   & J0928+6025$^a$ & J1001+5944$^a$ & J1009+0713$^b$ & J1435+3604$^a$ & J1553+3548$^a$  & J1616+4154     &  J1619+3342 \\ \hline
C  &  $>$-0.80      &  $>$-0.87      &  $>$-0.69      &  $>$-1.2       &  $>$-1.72      &  $>$-1.63      &  $>$-1.83      &  $>$-2.60      \\
N  & -0.63$\pm$0.16 & -1.08$\pm$0.17 & -1.50$\pm$0.20 & -1.40$\pm$0.14 & -1.03$\pm$0.17 &  $<$-1.64      &  $>$-1.08      & -1.74$\pm$0.16 \\
O  & -0.29$\pm$0.17 &  $>$-0.96      & -0.37$\pm$0.10 &  $>$-1.40      &  $>$-0.91      &  $>$-1.68      &  $>$-1.84      &  $>$-2.44      \\ 
Mg &  \nodata       &  $>$-0.89      &  \nodata       &$>$-0.86,$<$-0.21& \nodata       &  \nodata       &  \nodata       &  \nodata       \\
Si & -0.44$\pm$0.16 &  $>$-0.47      & -0.10$\pm$0.10 &  $>$-1.2       &  $>$-1.20      & -0.84$\pm$0.16 &  $>$-1.03      &  $>$-2.13      \\
P  &  $<$0.52       &  $<$1.39       &  0.06$\pm$0.12 &  $<$-0.31      &  $<$-0.44      &  \nodata       & -0.57$\pm$0.22 & -0.81$\pm$0.21 \\
S  &  $<$0.02       &  $<$0.15       &  $<$0.06       & -0.58$\pm$0.16 & -0.35$\pm$0.16 &  $<$-0.46      & -0.38$\pm$0.23 & -0.62$\pm$0.13 \\
Ti &  \nodata       &  $<$-0.32      &  \nodata       & -0.89$\pm$0.10 &  \nodata       &  \nodata       &  \nodata       & -1.56$\pm$0.11 \\
Fe & -0.90$\pm$0.19 &  0.10$\pm$0.17 & -0.47$\pm$0.11 & -0.84$\pm$0.20 & -0.83$\pm$0.15 & -0.99$\pm$0.17 & -1.03$\pm$0.21 & -1.62$\pm$0.18 \\
Ni &  $<$0.66       &  $<$0.12       &  $<$0.71       & -0.95$\pm$0.20 &  $<$-0.01      &  $<$0.17       &  \nodata       &  $<$-1.22 \\ \hline
\end{tabular}
\newline $^a$ Without applying photoionization correction factors. \hspace{2in} \newline $^b$ Measurements taken from \cite{Mei11}.
\end{table*}

\begin{table*}
\begin{center}
\caption{Ionization corrections, $\epsilon_{\rm{X}}$, for the sub-DLA absorbers \label{Tab:Cor}}
\begin{tabular}{cccccc}
\hline\hline 
X & J0925+4004 & J0928+6025 & J1001+5944 & J1435+3604 & J1553+3548 \\ 
 & -3.4$<$log $U$$<$-2.8 & -3.9$<$log $U$$<$-3.5 & -3.6$<$log $U$$<$-3.3 & log $U$$<$-3.3 & -3.8$<$log $U$$<$-2.7 \\ \hline
C  & -0.40,-0.22 & -0.22,-0.12 & -0.37,-0.27 & -0.16, 0.03 & -0.48,-0.15 \\
N  &  0.01, 0.03 &  0.01, 0.02 &  0.02, 0.03 &  0.00, 0.07 &  0.01, 0.04 \\
Mg & -0.30,-0.17 & -0.19,-0.10 & -0.30,-0.21 & -0.12, 0.04 & -0.33,-0.11 \\
Si & -0.45,-0.25 & -0.26,-0.14 & -0.39,-0.28 & -0.18, 0.00 & -0.38,-0.14 \\
P  & -0.37,-0.21 & -0.21,-0.12 & -0.37,-0.27 & -0.15,-0.01 & -0.48,-0.16 \\
S  & -0.25,-0.16 & -0.17,-0.11 & -0.28,-0.21 & -0.11, 0.00 & -0.30,-0.12 \\
Ti & -0.11,-0.10 & -0.13,-0.08 & -0.18,-0.15 & -0.06,-0.01 & -0.08,-0.07 \\
Fe & -0.16,-0.11 & -0.13,-0.08 & -0.19,-0.16 & -0.07, 0.00 & -0.14,-0.08 \\ 
Ni & -0.28,-0.21 & -0.22,-0.13 & -0.35,-0.28 & -0.15, 0.00 & -0.32,-0.17 \\ \hline
\end{tabular}
\newline \textbf{Note.} Values shown are the limits for $\epsilon_{\rm{X}}$ over the adopted range of log $U$.
\end{center}
\end{table*}

\begin{table}
\caption{Adopted Metallicity for each absorber \label{Tab:Metal}}
\begin{tabular}{ccccc}
\hline\hline 
  Absorber & $z_{abs}$ & log \nhI & Z & [Z/H] \\ 
           &          & (cm$^{-2}$)   &   &       \\ \hline
  J0925+4004 & 0.2477 & 19.55$\pm$0.15 & O & $-0.29\pm0.17$ \\
  J0928+6025 & 0.1538 & 19.35$\pm$0.15 & Fe& $+0.31\pm0.17$$^{ab}$ \\
  J1001+5944 & 0.3035 & 19.32$\pm$0.10 & O & $-0.37\pm0.10$ \\
  J1009+0713 & 0.1140 & 20.68$\pm$0.10 & S & $-0.58\pm0.16$ \\
  J1435+3604 & 0.2026 & 19.80$\pm$0.10 & S & $-0.41\pm0.16$$^a$ \\
  J1553+3548 & 0.0830 & 19.55$\pm$0.15 & Si& $-1.10\pm0.16$$^a$ \\
  J1616+4154 & 0.3211 & 20.60$\pm$0.20 & S & $-0.38\pm0.23$ \\
  J1619+3342 & 0.0963 & 20.55$\pm$0.10 & S & $-0.62\pm0.13$ \\ \hline
  \end{tabular}
\newline $^a$ After applying average $\epsilon_{\rm{X}}$ for element used. \newline $^b$ Applied a +0.3 dex correction, as discussed in \cite{Raf11}.
\end{table}

Since the column densities of sub-DLAs are not high enough to keep the H gas mainly neutral through self-shielding, it is necessary to correct for effects of photoionization through models. It has been suggested that corrections should be small for \nhI $>$ 19.5 at low-redshift \citep{Vie95,Jen09}. However, for completeness, we estimated corrections for all of the sub-DLAs. Ionization corrections for sub-DLA systems are determined using the \texttt{CLOUDY} ionization code (version 08.00, \citet{Fer98}). In this model, the gas is considered to be a uniform slab that is illuminated from one side by photoionizing radiation. We adopt the ionizing flux in the disk of the Milky Way based on the calculations of \citet{Fox05}. If sub-DLAs are associated with nearby galaxies, the flux from starlight in these galaxies should dominate over the extragalactic background in the soft UV region for distances as far as 100 kpc from the galaxy (see \citet{Fox05} for further details). \texttt{CLOUDY} models were also made using just the estimated extragalactic background flux from \citet{HM96} to determine if there would be significant differences. Even with the different shapes in the flux curves for these sources, the two models give almost identical results. Therefore, we feel it is reasonable to continue with our original approach using the galactic radiation.  

Oxygen is the best indicator of metallicity in sub-DLA systems because it traces the H$^0$ region due to charge exchange reactions between O and H. For this reason, its abundance was used as the initial estimate for metallicity in the \texttt{CLOUDY} models. If accurate O measurements were unavailable, we used S or Si. For the case of J0928+6025 we were forced to use Fe. Since Fe is prone to depletion, we will apply additional corrections to metallicity after modeling. Using these metallicities, \texttt{CLOUDY} models of the total column densities of elements as a function of ionization parameter, $U$, were made. 
The allowed range of $U$ is constrained using the measured values of the column densities for observed elements along with the observed ratio of Fe$^{++}$/Fe$^{+}$ and Si$^{++}$/Si$^{+}$. If the intermediate-ions are physically connected with the low-ions, as was discussed earlier, then these ratios should constrain the value of $U$ and also be less affected by dust depletion. An extensive study on ionization corrections in a multiphase ISM using \texttt{CLOUDY} by \citet{Mil10} found that using the Al$^{++}$/Al$^{+}$ ratio recovers the correct log $U$ value to within 0.1 dex for a two-phase medium.

For the majority of cases, the species N$^{0}$, Si$^{+}$, Ti$^{+}$, and Fe$^{+}$ appear to be underabundant relative to predicted \texttt{CLOUDY} abundances. The underabundance of Si, Ti, and Fe is likely due to their depletion onto dust grains. The underabundance of N will be discussed in the next section and is likely the result of its nucleosynthetic origins. It is also apparent that the measured abundances of the higher ions, such as Si$^{3+}$ and C$^{3+}$, do not appear to agree well with the lower ions in the \texttt{CLOUDY} models. However, as was mentioned earlier, the higher ions may be located in a different region than that of the lower ions and it would be unwise to use these states as a guide in determining the conditions of the lower ions. For several of our systems the kinematics of these high-ions do show significant difference from the low-ions, suggesting a multiphase gas component. The high-ions are still useful because the value of $U$ in those regions will certainly be higher than the low-ion regions and can be used as a conservative upper limit to the possible ionization.  

For J0925+4004, we used the upper limit of Fe$^{++}$/Fe$^{+}$ and the lower limit of Si$^{++}$/Si$^{+}$ to constrain log $U$ and this also agrees with the limit imposed from Si$^{3+}$. For J0928+6025, J1001+5944, J1435+3604 we use the Fe$^{++}$/Fe$^{+}$ ratio to constrain log $U$. For J1001+5944, this region agrees with the observed column densities of Fe$^{++}$, Si$^{+}$ and P$^{+}$. For J1553+3548, we used the upper limit suggested by the Si$^{3+}$ column density and the lower limit of Si$^{++}$/Si$^{+}$ to constrain $U$. The range of values we adopt for log $U$ can be seen in Table \ref{Tab:Cor}, along with the corresponding range in $\epsilon_{\rm{X}}$. We will adopt the average value of $\epsilon_{\rm{X}}$ for each individual element when determining the metallicity for these systems. For example, in J1553+3548 we find $-0.38<\epsilon_{\rm{Si}}<-0.14$ and so the correction we will use for this system's metallicity, which is determined from Si in this case, is $\epsilon_{\rm{Si}}=-0.26$. Since O has negligible ionization correction because of its relationship to H, the metallicity measurements for sub-DLAs J0925+4004 and J1001+5944 are independent from the degree of photoionization. 

The metallicities adopted for the DLAs J1616+4154 and J1619+3342 were based on the abundance derived from the S~II lines because this ion provides accurate column density measurements and it has relatively low depletion rate (however, see the caveats discussed by \citet{Jen09}). For both cases, this value can be compared to the abundance found for P, which also has low depletion levels. For J1616+4154 and J1619+3342, we found [S/P]=$0.19\pm0.16$ and [S/P]=$0.19\pm0.21$, respectively. The metallicities found from either of these elements appears consistent with the other within the uncertainties. However, it may suggest $\alpha$ enhancement. The metallicities we adopt for each of our absorbers is shown in Table \ref{Tab:Metal}.

\begin{figure*}
\plotone{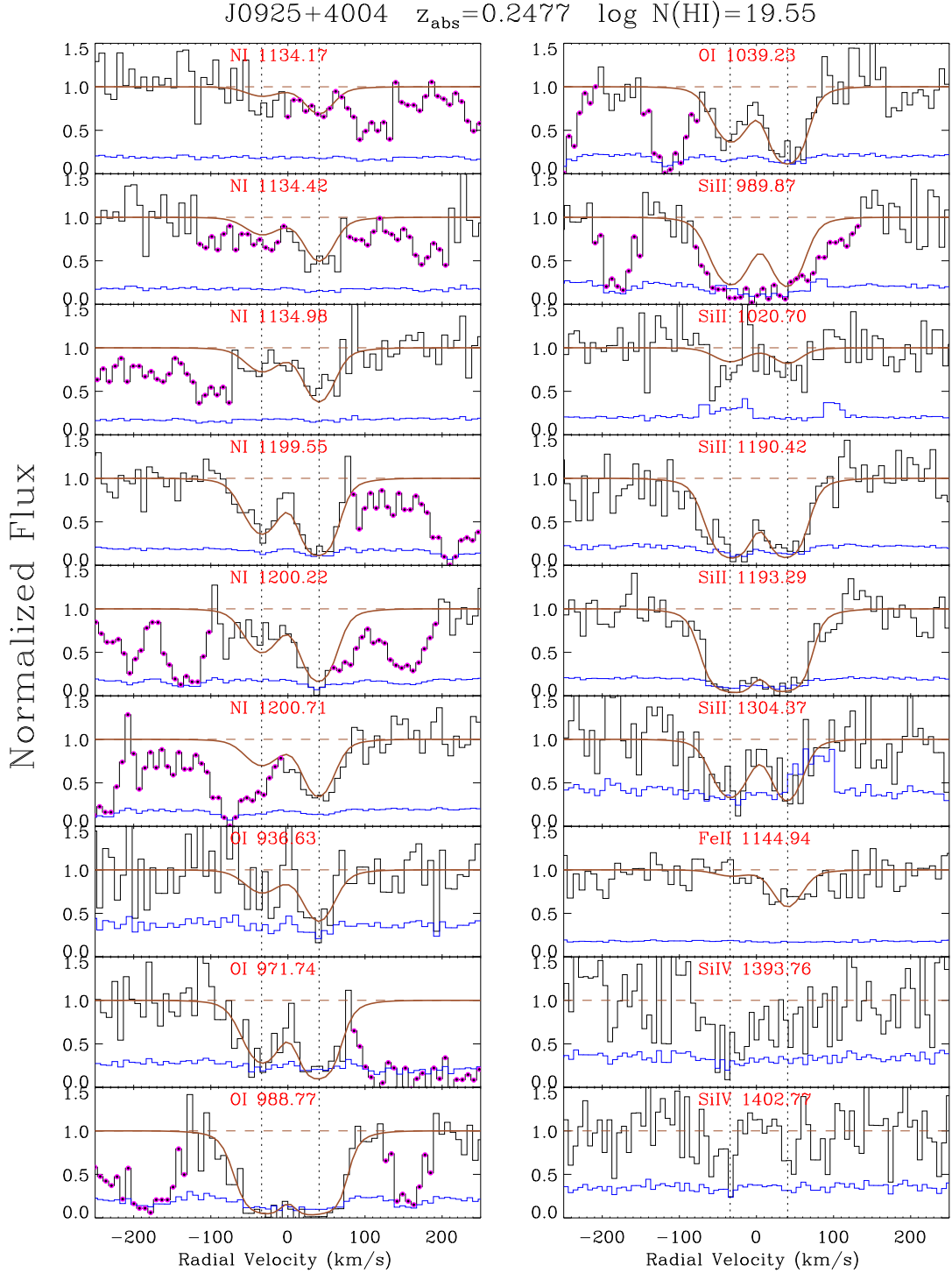}
\caption{Velocity plots for several lines in the COS spectra of J0925+4004. The dashed lines indicate the velocity components, the data is presented in black, the error array is given in blue, and the best fit is brown. The Fe~II \lam 1144.94 line required a large offset to fit ($-40$ \kms) suggesting that it might be blended with something. The following lines were not used for profile fits: N~I \lala 1200.22 and 1200.71, since they are significantly blended together; Si~II \lam 989.87 appears blended, perhaps with N~III \lam 989.80. Unrelated features or blends are indicated with magenta circles. Rest velocity corresponds to $z=0.2477$. \label{Fig:J0925vel}}
\end{figure*}

\begin{figure}
\includegraphics[width=3.5in]{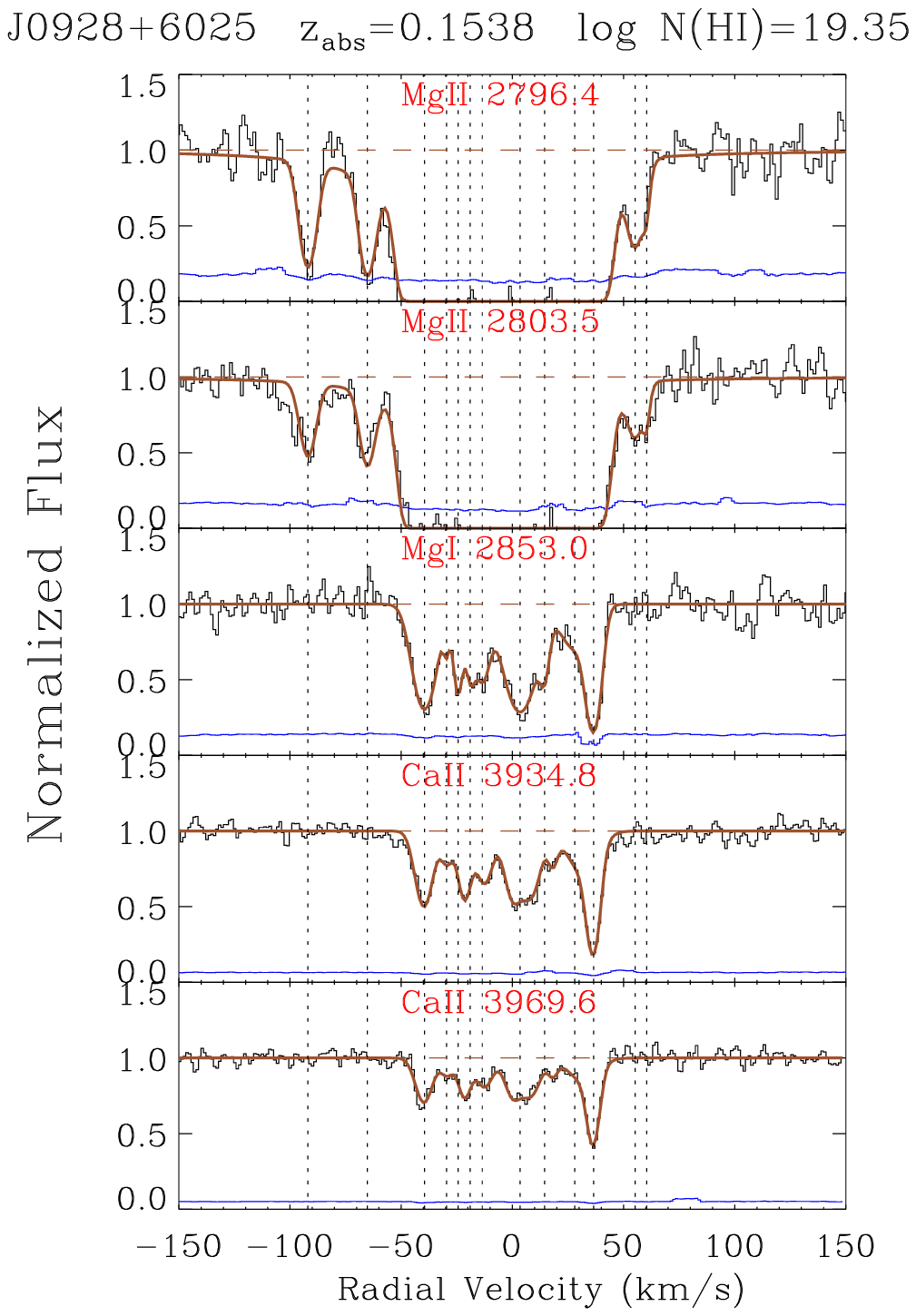}
\caption{Velocity plots for several lines in the HIRES spectra of J0928+6025. The dashed lines indicate the velocity components, the data is presented in black, the error array is given in blue, and the best fit is brown. Rest velocity corresponds to $z=0.1538$. \label{Fig:J0928keckvel}}
\end{figure}

\begin{figure*}
\plotone{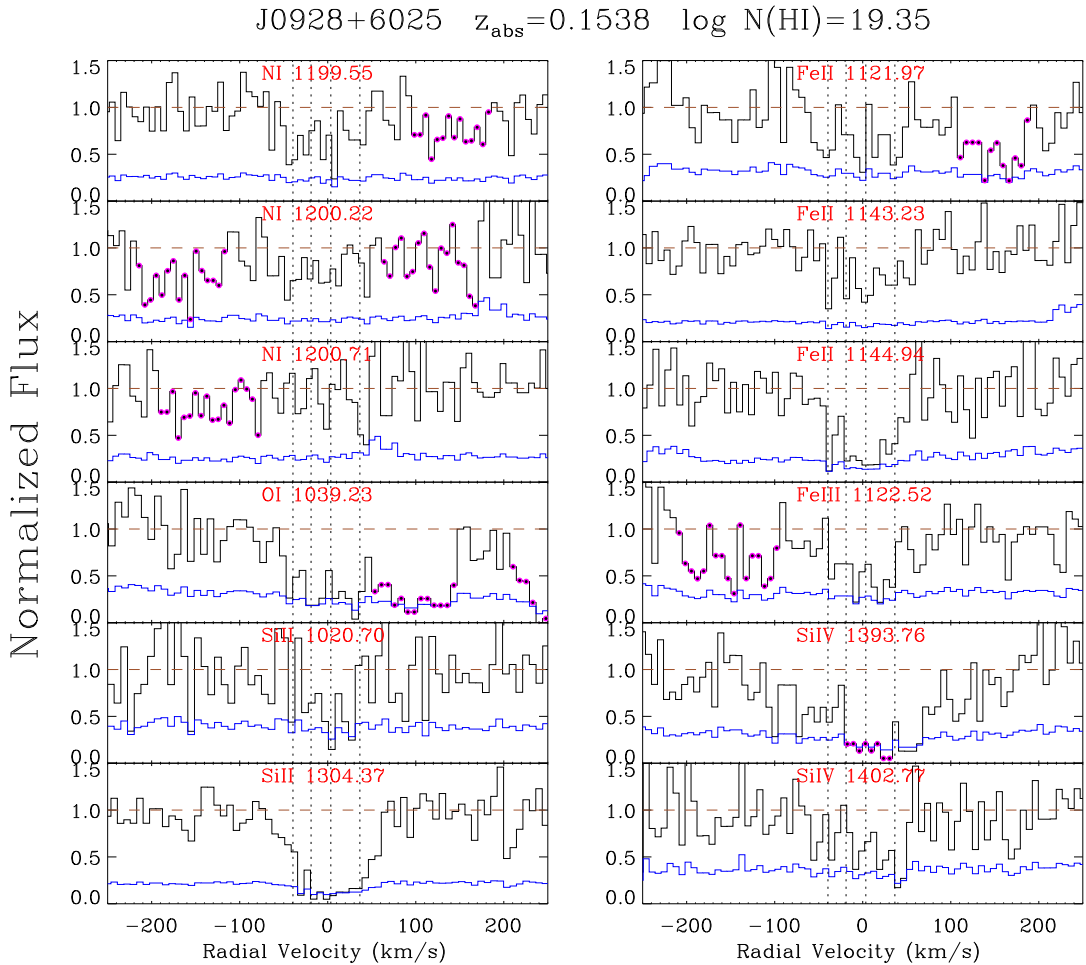}
\caption{Velocity plots for several lines in the COS spectra of J0928+6025. The dashed lines indicate the velocity components, the data is presented in black, the error array is given in blue, and the best fit is brown. Unrelated features or blends are indicated with magenta circles. O~I \lam 1039.23 is blended with N~I \lam 1199.55 in the MW. Si~IV \lam 1393.76 is blended with Fe~II \lam 1608.45 line in the Milky Way (MW). Rest velocity corresponds to $z=0.1538$. Fits were not made for this system because of the low S/N. \label{Fig:J0928vel}}
\end{figure*}

\begin{figure*}
\plotone{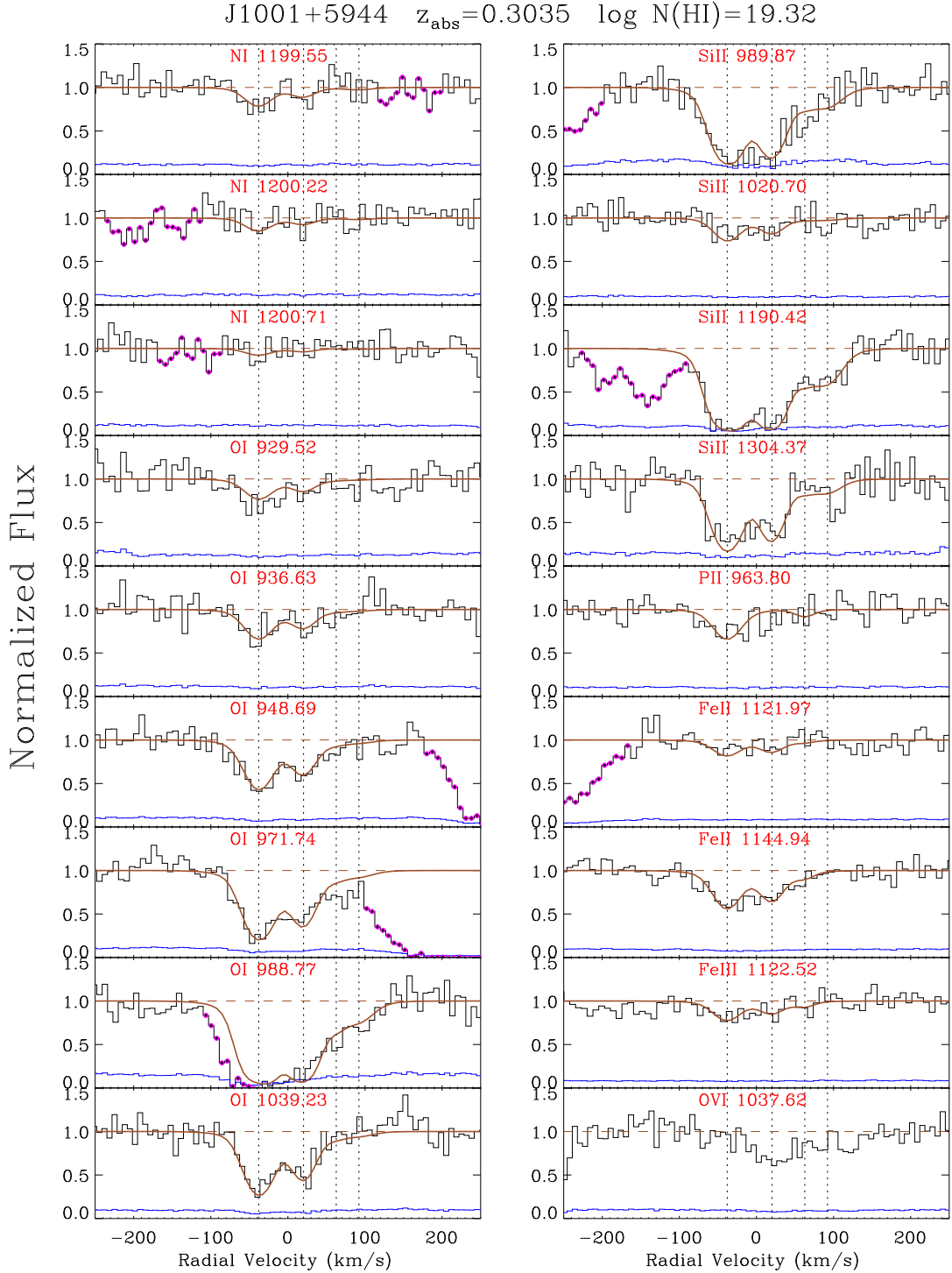}
\caption{Velocity plots for several lines in the COS spectra of J1001+5944. The dashed lines indicate the velocity components, the data is presented in black, the error array is given in blue, and the best fit is brown. Unrelated features or blends are indicated with magenta circles. The following lines were not used for profile fits: O~I \lam 971.74 is blended with the Ly$\gamma$ line of the same system; O~I \lam 988.77 appears to be blended with and unidentified line based on its strength. Rest velocity corresponds to $z=0.3035$. \label{Fig:J1001vel}}
\end{figure*}

\begin{figure*}
\plotone{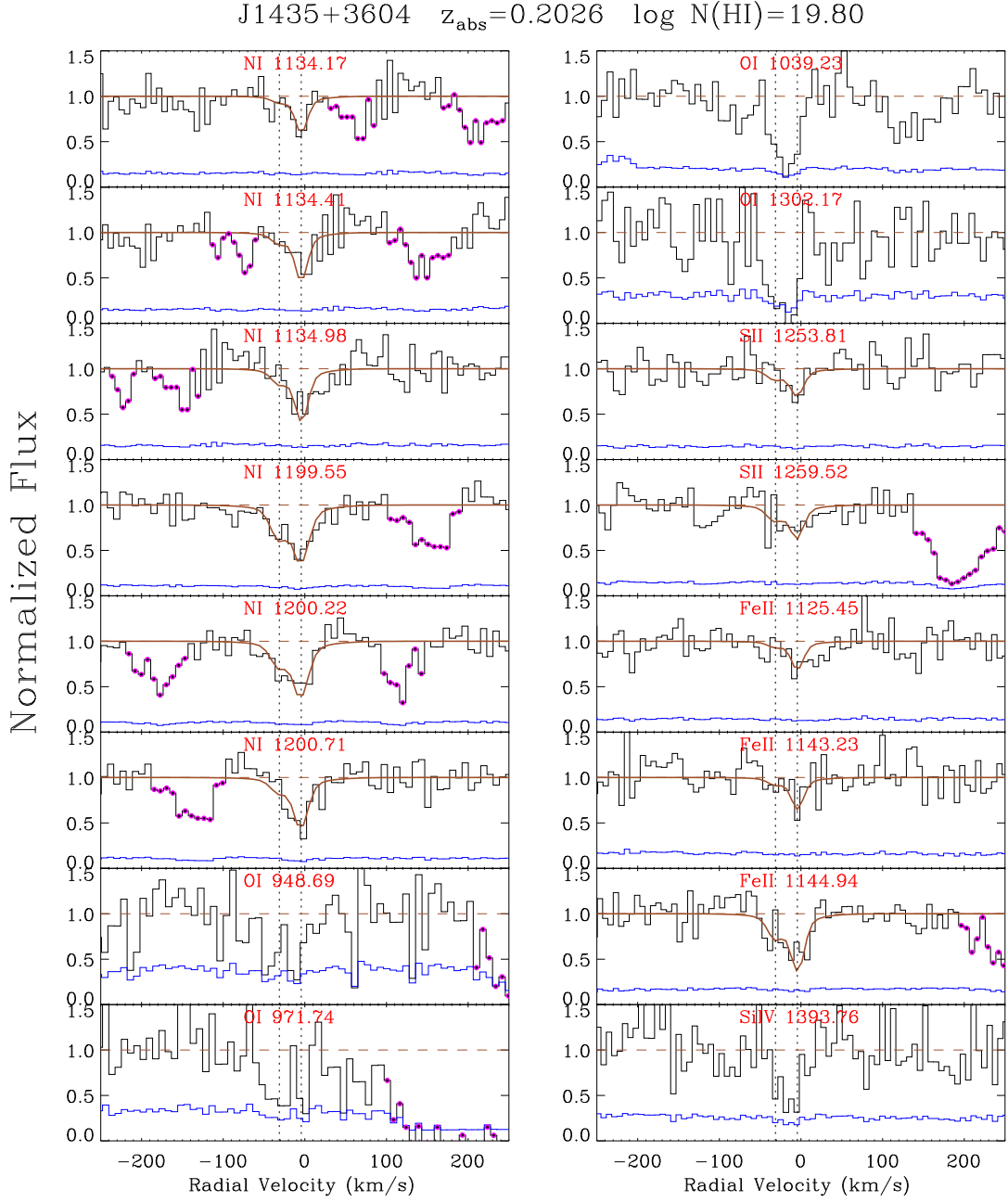}
\caption{Velocity plots for several lines in the COS spectra of J1435+3604. The dashed lines indicate the velocity components, the data is presented in black, the error array is given in blue, and the best fit is brown. Unrelated features or blends are indicated with magenta circles. Si~II \lam 1020.70 is blended with the Ly$\gamma$ line of an absorber at $z_{abs}=0.2624$. Si~IV \lam 1402.77 is blended with the Ly$\alpha$ line of an absorber at $z_{abs}=0.3874$. Rest velocity corresponds to $z=0.2027$. \label{Fig:J1435vel}}
\end{figure*}

\begin{figure*}
\plotone{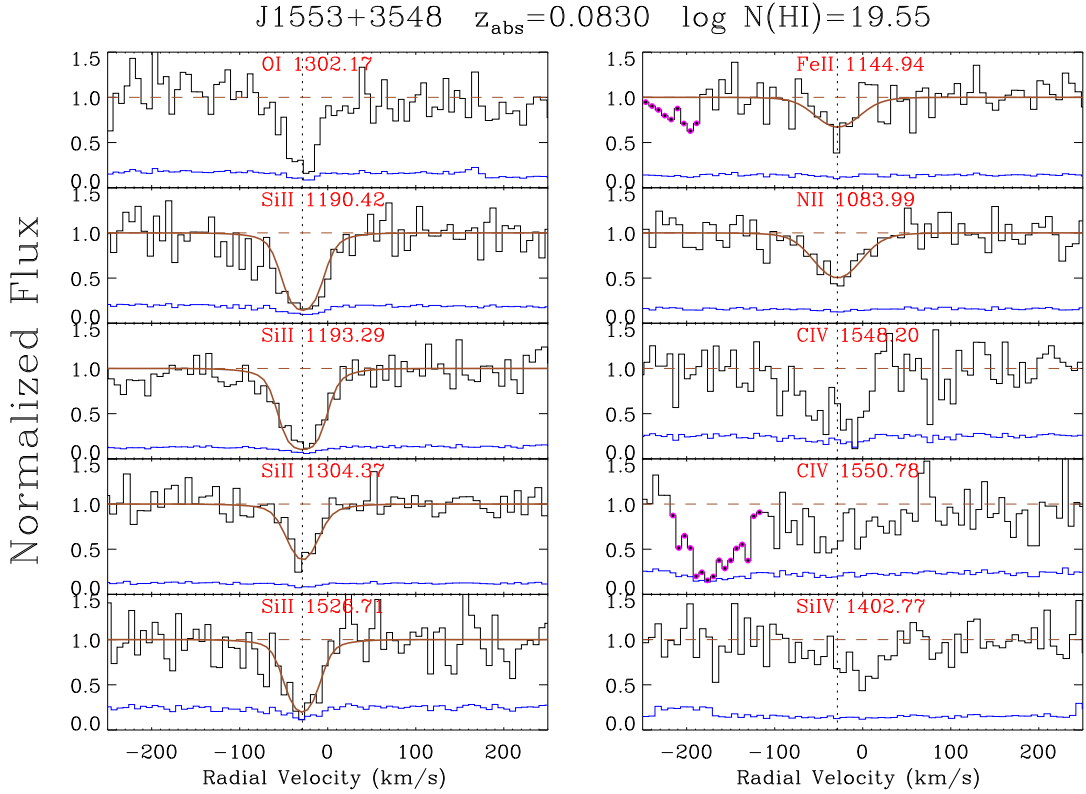}
\caption{Velocity plots for several lines in the COS spectra of J1553+3548. The dashed lines indicate the velocity components, the data is presented in black, the error array is given in blue, and the best fit is brown. Unrelated features or blends are indicated with magenta circles. The Si~IV \lam 1393.76 appears to be blended with an unidentified line. Rest velocity corresponds to $z=0.0830$. \label{Fig:J1553vel}}
\end{figure*}

\begin{figure*}
\plotone{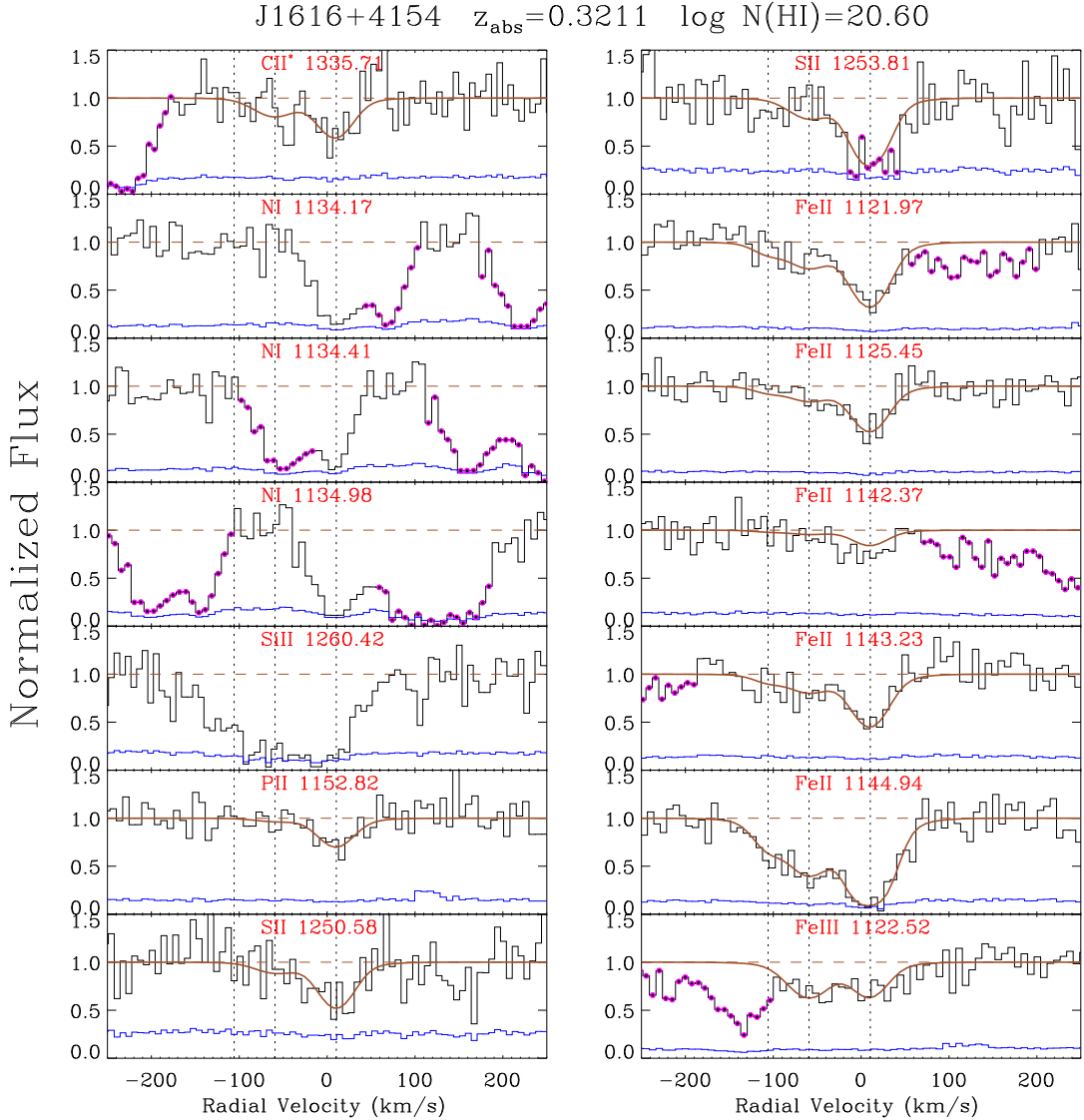}
\caption{Velocity plots for several lines in the COS spectra of J1616+4154. The dashed lines indicate the velocity components, the data is presented in black, the error array is given in blue, and the best fit is brown. Unrelated features or blends are indicated with magenta circles. The N~I lines show strong blending with each other. S~II \lam 1253.81 is blended with C~I \lam 1656.93 in the MW. S~II \lam 1259.52 is blended with Si~II \lam 1260.42. Rest velocity corresponds to $z=0.3213$. \label{Fig:J1616vel}}
\end{figure*}

\begin{figure}
\includegraphics[width=3.5in]{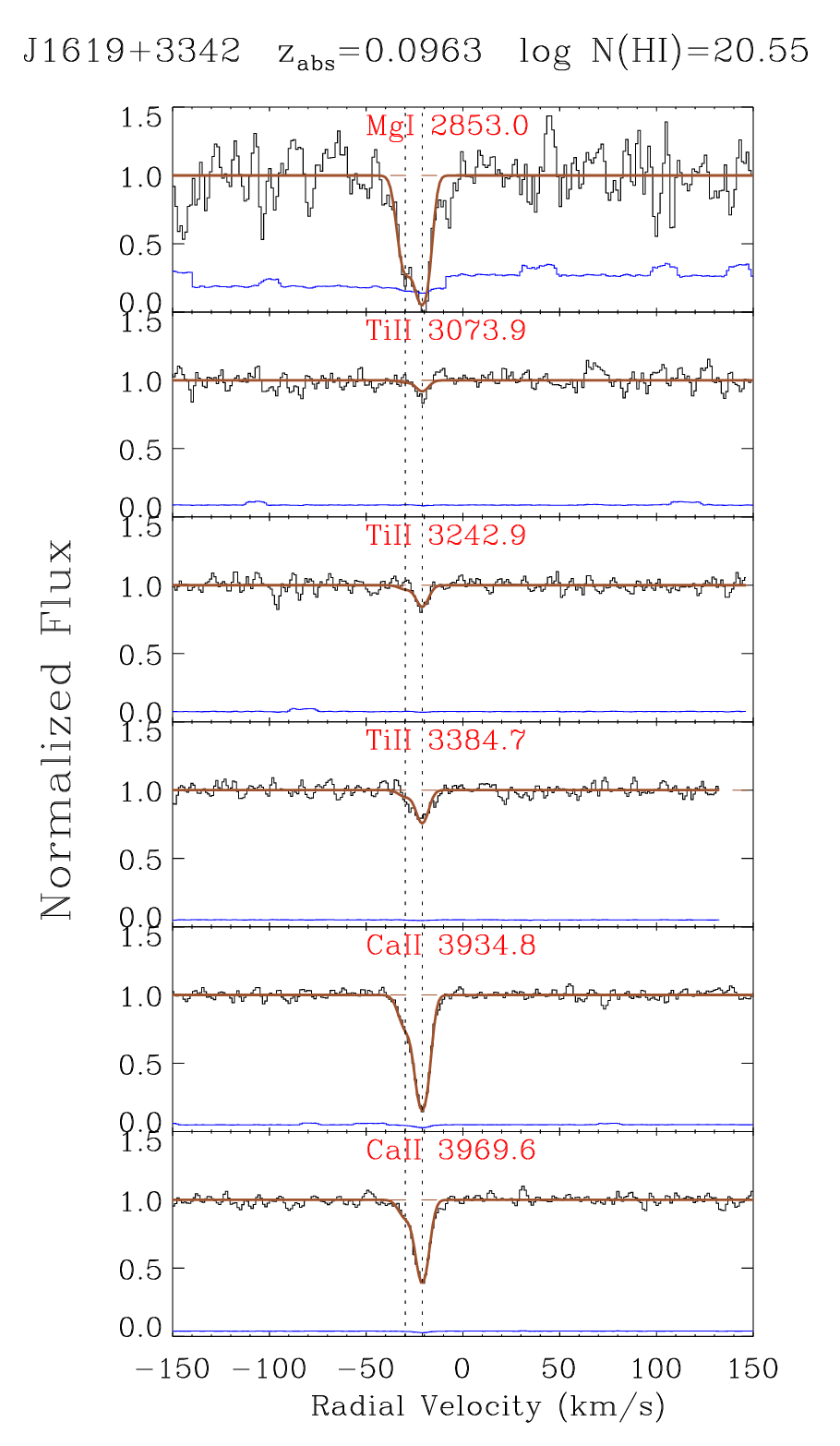}
\caption{Velocity plots for several lines in the HIRES spectra of J1619+3342. The dashed lines indicate the velocity components, the data is presented in black, the error array is given in blue, and the best fit is brown. Rest velocity corresponds to $z=0.0964$. \label{Fig:J1619keckvel}}
\end{figure}

\begin{figure*}
\plotone{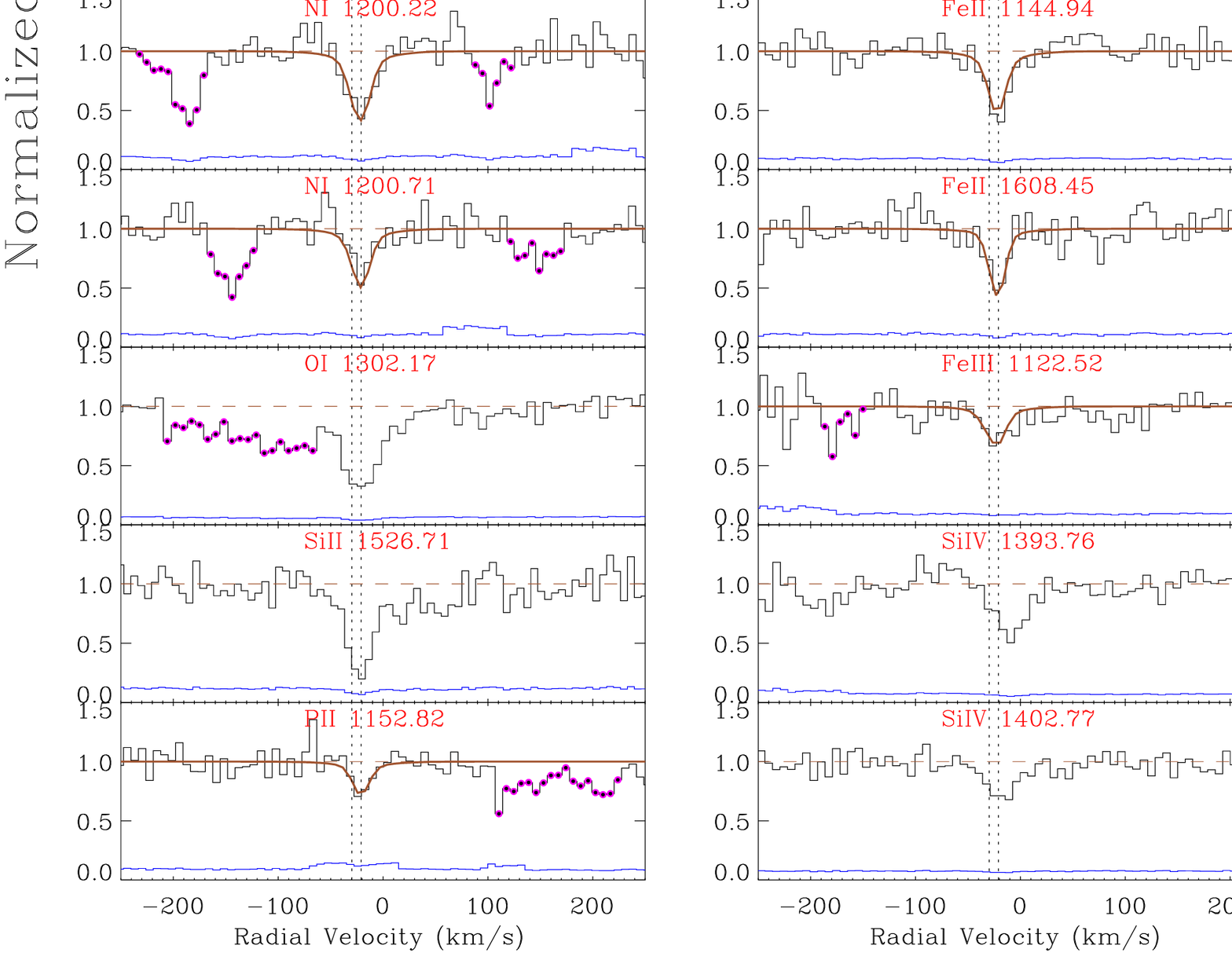}
\caption{Velocity plots for several lines in the COS spectra of J1619+3342. The dashed lines indicate the velocity components, the data is presented in black, the error array is given in blue, and the best fit is brown. Unrelated features or blends are indicated with magenta circles. Fe~II \lam 1143.23 is blended with S~II \lam 1253.81 in the MW. Rest velocity corresponds to $z=0.0964$. \label{Fig:J1619vel}}
\end{figure*}

\subsection{Depletion Patterns}
The depletion patterns for these systems were estimated using the model of depletion patterns in the ISM of the Milky Way from \citet{Jen09}. In this model a single value (F$_{*}$) indicates the overall level of depletion in a sightline, with larger values of F$_{*}$ indicating higher levels of dust depletion. Accurate determination of F$_{*}$ requires a large number of elements which are spread over a broad range of condensation temperatures. Only a fraction of these lines are available to us because of the low S/N of our data and limited wavelength coverage. Despite this, it is still possible to see general trends the depletion patterns of these systems. 

We are able to estimate a depletion factor of F$_{*}$ for J0925+4004, J1001+5944, J1435+3548 and J1619+3342, as these systems have a sufficient number of lines for basic analysis. For J0925+4004 we find F$_{*}=-0.13\pm0.19$, implying an intrinsic metallicity of [Z/H]$=-0.41\pm0.15$, which is consistent with the metallicity determined from the O~I lines, [O/H]$=-0.29\pm0.17$. For J1001+5944 we find F$_{*}=-0.69\pm0.15$, implying an intrinsic metallicity of [Z/H]$=-0.79\pm0.13$, which is below the value determined from the O~I line, [O/H]$=-0.37\pm0.10$. The depletion pattern for J1001+5944 does not seem to fit the linear trend of F$_{*}$ very well, since there is a small number of diagnostic lines, this is likely the source of discrepancy. The deficiency of N in this system may also contribute to the problem. For J1435+3548 we find F$_{*}=-0.78\pm0.22$, implying an intrinsic metallicity of [Z/H]$=-0.97\pm0.21$, which is below the value determined from the S~II line, [S/H]$=-0.41\pm0.16$. For J1619+3342 we find F$_{*}=-0.55\pm0.11$, implying an intrinsic metallicity of [Z/H]$=-1.53\pm0.17$, which is well below the metallicity determined from the S~II and P~II lines, [S/H]$=-0.62\pm0.13$. For the latter two systems, one possible reason for the discrepancy is that both P and S show somewhat anomalous behavior in the way they deplete (i.e. \citet{Jen09}). Additionally, some of the S~II could arise
from regions where the hydrogen is ionized, since the ionization potential of S~II is 23.4 eV, which could create a false impression that the abundance of S relative to H is higher than reality. All of our systems show a trend towards negative values of F$_{*}$ which indicates little, if any, depletion. This is typically seen along sightlines with extremely low average densities in the halo of our Galaxy \citep{Jen09}, implying that each of these sightlines pass through the outer edge of a galaxy.

\subsection{Metallicity Evolution of DLAs and Sub-DLAs}
The absorption systems in this study show a range of metallicities from above solar for J0928+0625, to one-tenth solar for J1553+3548. We now compare the observed metallicities of our low-redshift DLAs and sub-DLAs to those in the literature. The nearly undepleted element Zn is a commonly adopted metallicity tracer in DLA and sub-DLA systems and is used for some of the measurements in the sample we constructed (However, see see the caveats noted by \citet{HS99}). Most of the other measurements are based on abundances of $\alpha$-elements (O, Si, S). The use of $\alpha$-elements with Zn measurements does create an inhomogeneous sample because of the different nucleosynthetic origins between these elements, however \citet{Kul07} found that their inclusion in the samples made essentially no change in their results. The findings by \citet{PW02} that [Si/Zn]$=0.03\pm0.05$ also suggests that combining these populations seems reasonable. The remaining sample is made up of metallicities determined from Fe using a +0.3 dex correction for depletion.

\begin{figure*}
\plotone{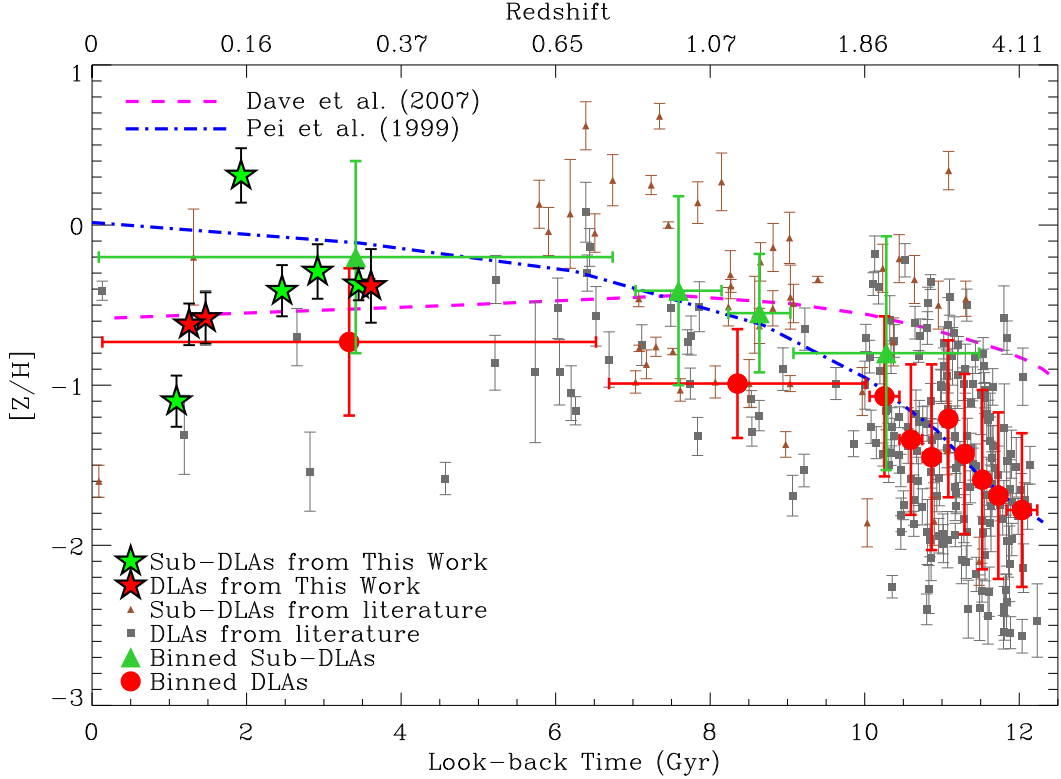}
\caption{[Z/H] vs look-back time for our low-redshift systems along with those from the literature. The red circles and green triangles correspond to unweighted mean metallicity of each bin for the DLAs and sub-DLAs, respectively. Horizontal bars denote the range in look-back times spanned by each bin. \label{Fig:lookback}}
\end{figure*}

The comparison sample we will utilize includes 194 DLAs with $0.0096<z<4.743$ and 53 sub-DLAs with $0.0063<z<3.1710$. The majority of DLA sources come from the list constructed by \citet{Raf11}. This list is designed to be unbiased at $z>1.5$. The majority of sub-DLA sources comes from \citet{Kul07} and \citet{Mei09}. The DLA sample was divided into 10 redshift bins with equal numbers of sources in each bin. The unweighted and weighted mean metallicity was determined for each bin and is shown in Table \ref{Tab:meanDLA}. Weighted metallicities are defined as:
\begin{equation}
[Z/H]_{wm}=\frac{\Sigma w_i [Z/H]_i}{\Sigma w_i}
\end{equation}
Where the weight, $w_i$, is the reciprocal square of the corresponding uncertainty ($w_i=1/\sigma^2_i$). This was repeated for the sub-DLAs over 4 redshift bins and these are shown in Table \ref{Tab:meansubDLA}. The unweighted binned values can be seen alongside chemical evolution models from \citet{Dav07} (magenta dashed line) and \citet{Pei99} (blue dash-dot line) in Figure \ref{Fig:lookback}. The look-back times are estimated assuming a cosmological model with $\Omega_m=0.30$, $\Omega_{\Lambda}=0.70,$ and $H_0=70$ \kms\ Mpc$^{-1}$. For each bin, the data points are plotted at the median value of the bin and the horizontal error bars denote the range in look-back time spanned by that bin. The vertical error bars denote the uncertainty in the mean metallicity and accounts for scatter and also measurement uncertainties. The model from \citet{Dav07} is a hydrodynamic simulation whereas the model from \citet{Pei99} model is semi-analytical. The \citet{Pei99} model agrees with the DLA systems at $z_{abs}>1.8$, but overpredicts the mean metallicity of low-redshift systems. The model from \citet{Dav07} only agrees with low-redshift DLA observations, and seems to overpredict the mean metallicities in the highest redshift systems. The binned sub-DLA data agree with both models, however there is a deficiency in data points at high-redshifts.

As expected, the low-redshift systems do appear to have higher average metallicities than higher redshift systems. We also see a relatively large scatter in metallicity present among our sources as we would expect from random probing of galaxies without regard to mass. As was the case in previous studies \citep{Kul07,Pro03b}, we continue to see an underabundance in mean metallicity at low-redshift relative to solar. This trend shows agreement with recent cosmological hydrodynamic simulations \citep{Dav07,Cen10}. It is also apparent that sub-DLAs, on average, show slightly higher mean metallicities than DLAs, but not to extent found previously by \citet{Kul07}. However, the reasons for this trend are still unclear. 


\begin{table}
\begin{center}
\caption{Binned DLA Mean Metallicities \label{Tab:meanDLA}}
\begin{tabular}{cccc}
\hline\hline 
Redshift Bin & \# of systems & Unweighted & Weighted \\ \hline
0.0096-0.7405 & 19 & -0.73$\pm$0.46 & -0.68$\pm$0.46 \\
0.7731-1.8644 & 19 & -0.99$\pm$0.33 & -1.02$\pm$0.33 \\
1.8918-2.1408 & 19 & -1.07$\pm$0.50 & -1.13$\pm$0.50 \\
2.1410-2.3700 & 19 & -1.34$\pm$0.47 & -1.20$\pm$0.47 \\
2.3745-2.5841 & 19 & -1.45$\pm$0.58 & -1.18$\pm$0.58 \\
2.5950-2.7835 & 19 & -1.21$\pm$0.49 & -1.23$\pm$0.49 \\
2.7958-3.0621 & 20 & -1.44$\pm$0.50 & -1.26$\pm$0.50 \\
3.1040-3.3655 & 20 & -1.59$\pm$0.56 & -1.57$\pm$0.56 \\
3.3869-3.7609 & 20 & -1.69$\pm$0.52 & -1.75$\pm$0.52 \\
3.7612-4.7430 & 20 & -1.78$\pm$0.48 & -1.78$\pm$0.48 \\ \hline
\end{tabular}
\end{center}
\end{table}

\begin{table}
\begin{center}
\caption{Binned sub-DLA Mean Metallicities \label{Tab:meansubDLA}}
\begin{tabular}{cccc}
\hline\hline 
Redshift Bin & \# of systems & Unweighted & Weighted \\ \hline
0.0063-0.7821 & 13 & -0.20$\pm$0.60 & -0.36$\pm$0.60 \\
0.8426-1.1157 & 13 & -0.41$\pm$0.59 & -0.51$\pm$0.59 \\
1.1414-1.4094 & 13 & -0.55$\pm$0.37 & -0.64$\pm$0.37 \\
1.4244-3.1710 & 14 & -0.80$\pm$0.73 & -0.43$\pm$0.73 \\ \hline
\end{tabular}
\end{center}
\end{table}


\section{Nitrogen Nucleosynthesis}
The nucleosynthetic origins of nitrogen is a topic that has proved to be quite complicated. The production of N occurs from primary and secondary production processes in intermediate mass stars. Primary production occurs from newly synthesized C via the CNO cycle and secondary production uses C and O from previous star formation in the CNO cycle \citep{Mar01}. Measurements of high-redshift DLAs, where N~I lines are redshifted into wavelengths where ground-based observing is possible, find N to be underabundant relative to $\alpha$-elements \citep{Cen98,Hen07,Pet08}. The study of N relative to $\alpha$-process elements is useful because it can give insights into the star formation history of a gas system, since [N/$\alpha$] will increase as a system ages and intermediate mass stars have time to inject more N into the system. 

The $\alpha$-process elements that we will be utilizing are O, Si, and S. The specific $\alpha$-element used was determined for each system in the following way. The primary choice was O, because of its resonant exchange with H. The next choice for $\alpha$ was S, since it shows only mild depletion onto dust grains. If neither O or S are accurately determined, then the abundance for Si is adopted. However, since Si shows a strong tendency to deplete onto dust grains this estimate is less certain. The measured values for [N/$\alpha$] are listed in Table \ref{Tab:Nalph}. It appears to be the case for the majority of our absorption systems that N is underabundant relative to $\alpha$-process elements.

\begin{table}
\begin{center}
\caption{[N/$\alpha$] and [$\alpha$/H] for each absorber \label{Tab:Nalph}}
\begin{tabular}{cccc}
\hline\hline 
ID  & $\alpha$ & $[\rm{N}/\alpha]$ & $[\alpha/\rm{H}]$  \\ \hline
J0925+4004 & O & -0.32$\pm$0.10$^a$ & -0.29$\pm$0.17 \\ 
J0928+6025 & O &  $<$-0.10$^a$     &  $>$-0.67 \\ 
J1001+5944 & O & -1.10$\pm$0.17$^a$ & -0.37$\pm$0.10 \\ 
J1009+0713 & S & -0.82$\pm$0.16 & -0.58$\pm$0.16 \\ 
J1435+3604 & S & -0.58$\pm$0.18$^a$ & -0.41$\pm$0.16$^a$ \\
J1553+3548 & Si&  $<$-0.51$^a$      & -1.10$\pm$0.16$^a$ \\ 
J1616+4154 & S &  $>$-0.70      & -0.38$\pm$0.23 \\  
J1619+3342 & S & -1.12$\pm$0.15 & -0.62$\pm$0.13 \\ \hline
\end{tabular}
\newline $^a$ Applied average $\epsilon_{\rm{X}}$ for elements used.
\end{center}
\end{table}

\begin{figure*}
\plotone{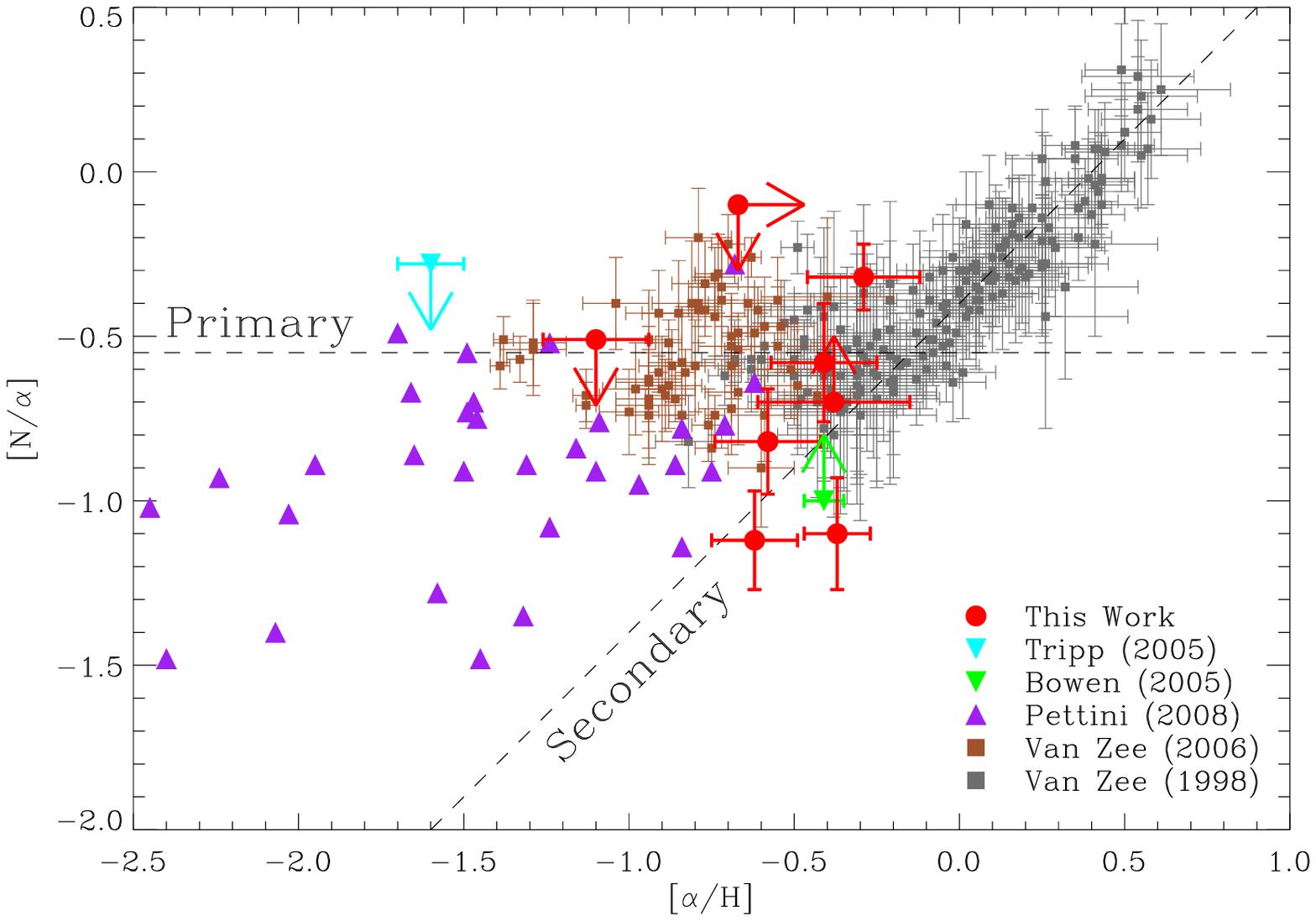}
\caption{[N/$\alpha$] vs [$\alpha$/H] for the 8 systems discussed in this paper (red points), high-redshift DLAs from \citet{Pet08}, H~II regions of nearby dwarf galaxies from \citet{Van06}, the low-redshift sub-DLA system from \citet{Tri05}, the low-redshift DLA from \citet{Bow05}, and H~II regions in nearby spiral galaxies from \citet{Van98}. Correction factors have been applied here for the sub-DLAs. The dashed lines are approximate representations of primary and secondary levels of N production. \label{Fig:Nitrogen}}
\end{figure*}

Abundance measurements of N have been measured previously using optical spectra of H~II regions of nearby low-luminosity galaxies, where N is mainly produced from primary production, by \citet{Van06} and show a flat plateau in [N/O] ratio with respect to metallicity. This ratio was also measured in higher metallicity H~II regions in nearby spiral galaxies, where more secondary production occurs, by \citet{Van98} and these show a steady increase in [N/O] for larger metallicities. In Figure \ref{Fig:Nitrogen}, we show [N/$\alpha$] vs. [$\alpha$/H] for our sample along with these H~II regions, as well as the high-redshift DLA sample from \citet{Pet08}, the low-redshift sub-DLA system from \citet{Tri05}, and a low-redshift DLA from \citet{Bow05}. \citet{Pet08} suggested the large discrepancy in high-redshift DLAs could be due to delayed enrichment of N with respect to O, estimated by \citet{Hen00} to be $\sim$250 Myr, which is a more considerable fraction of time at high-redshift (see \citet{Pet08} for a more detailed discussion). Interestingly, all of our low-redshift systems also lie to left the transition point of secondary production seen in the H~II regions of local spiral galaxies. The large spread in [N/$\alpha$] is similar to that seen in high-redshift systems. This result seems puzzling if we expect these systems to have had sufficient time to begin secondary production. Despite this, the $\alpha$ abundances of the low-redshift DLAs and sub-DLAs are higher than that of the high-redshift systems, as one would expect.

\section{Affiliated Galaxies}
The lack of information about the host galaxies of DLAs and sub-DLAs has been an obstacle that has limited the exploitation of these systems for probing galaxy evolution. The deficiency in the number of low-redshift systems discovered to date has only hampered these efforts more. Some progress has been made recently using photometric redshifts and colors to identify affiliated galaxies for 80 Mg~II selected absorption systems at redshift $0.1\le z_{abs} \le1.0$ \citep{Rao11}. However, our sample offers a unique opportunity to identify affiliated galaxies for systems not biased on Mg~II absorption.

\begin{figure*}
\begin{center}$
\begin{array}{cc}
\includegraphics[height=0.48\linewidth,angle=90]{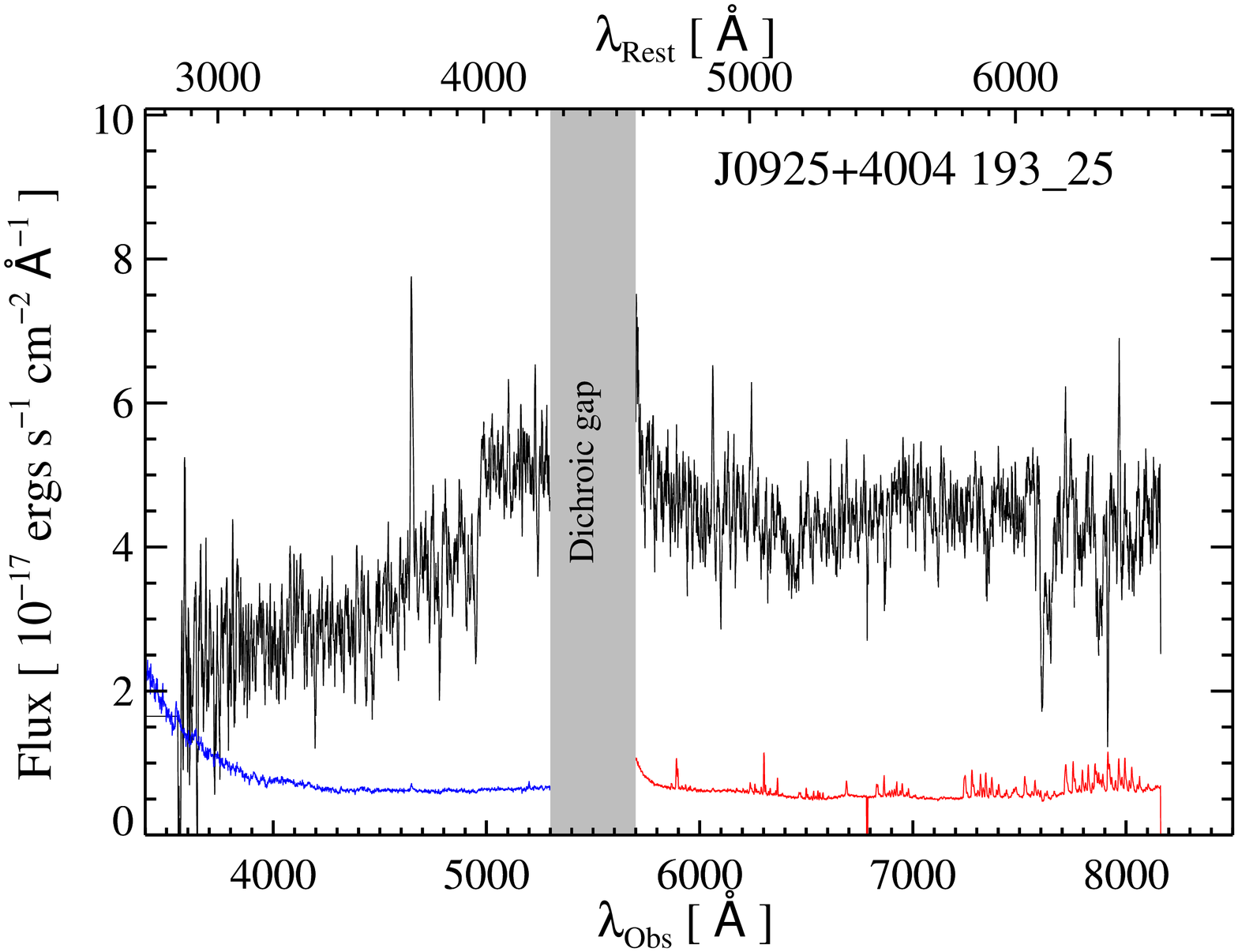} &
\includegraphics[height=0.48\linewidth,angle=90]{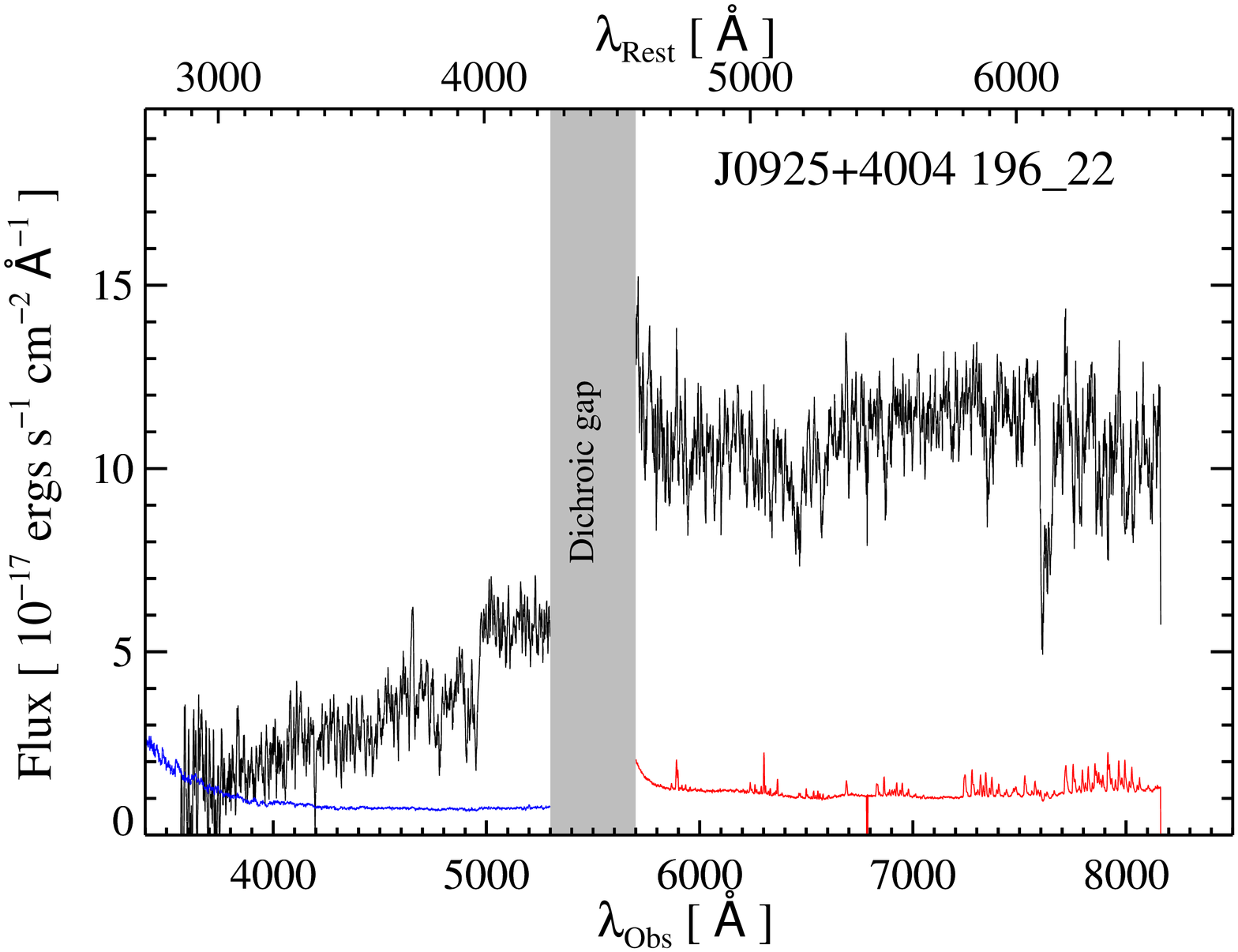} \\
\includegraphics[height=0.48\linewidth,angle=90]{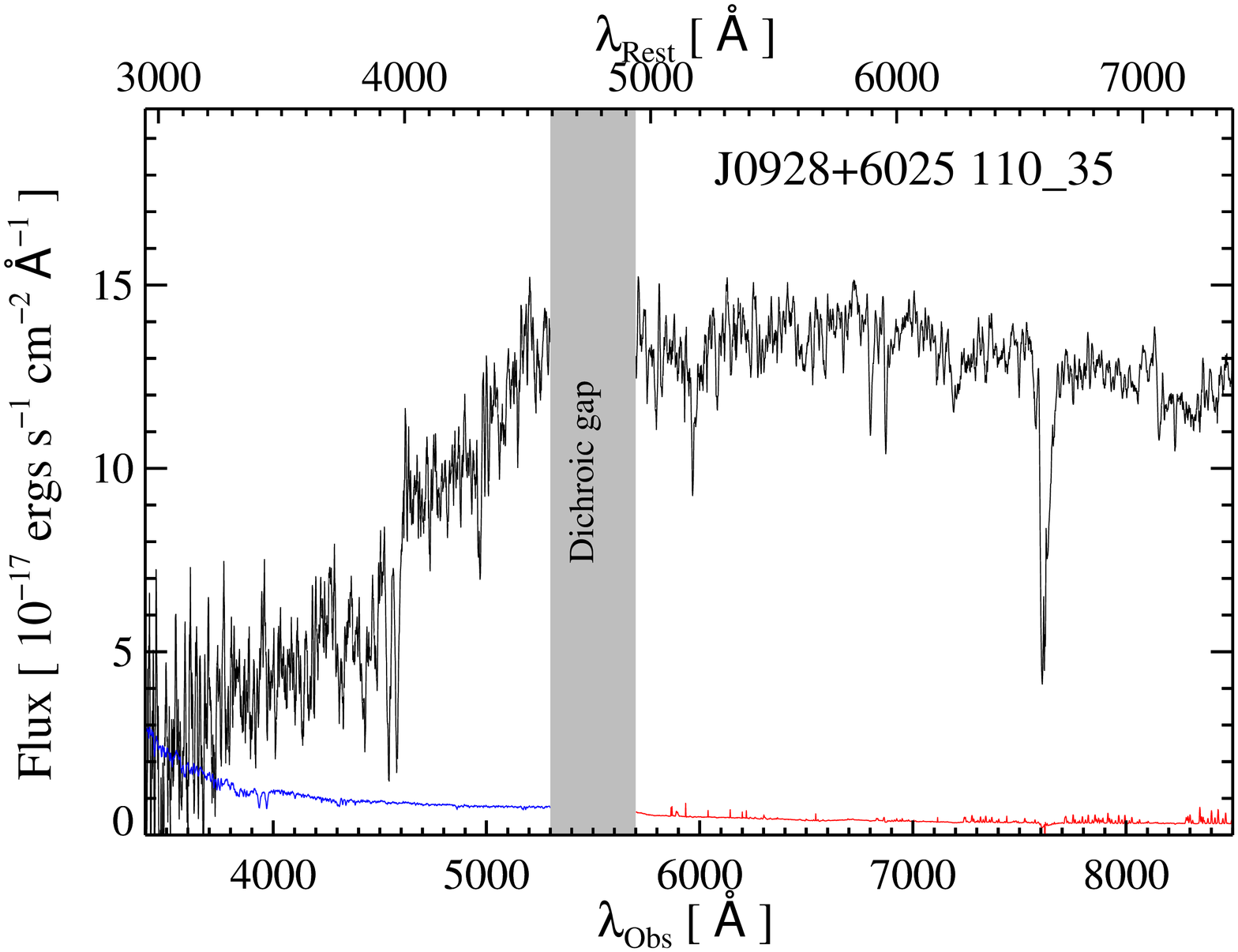} &
\includegraphics[height=0.48\linewidth,angle=90]{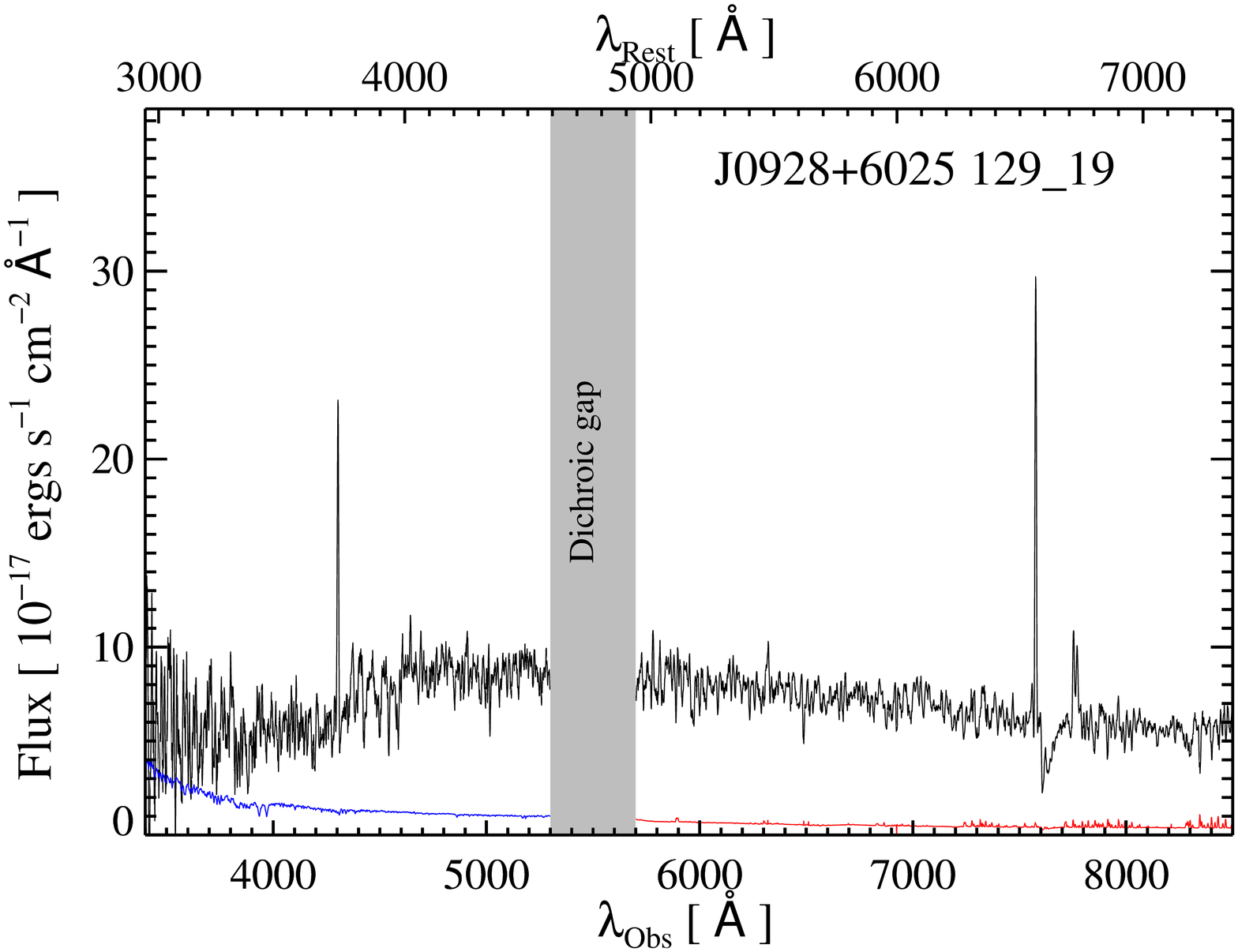} \\
\includegraphics[height=0.48\linewidth,angle=90]{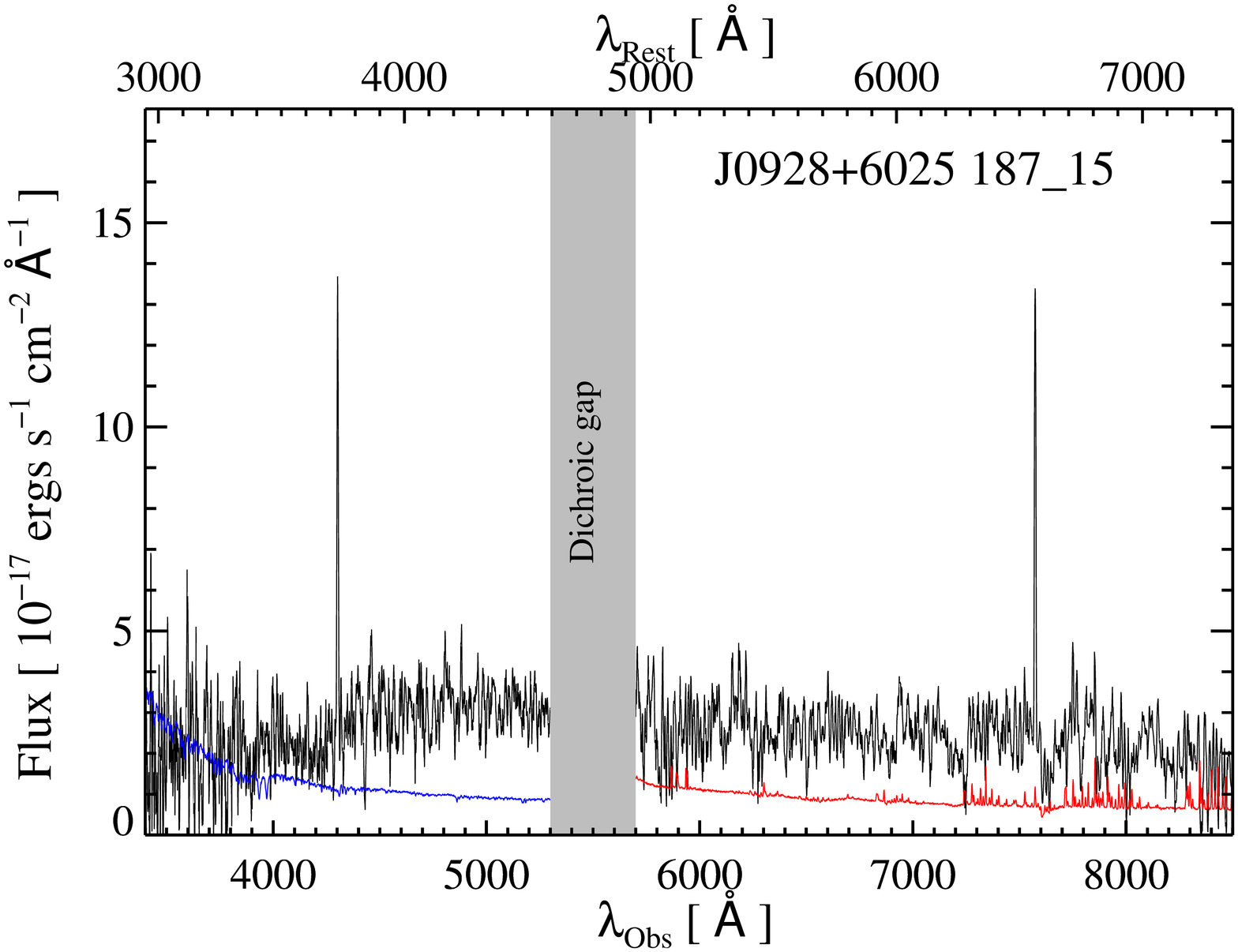} &
\includegraphics[height=0.48\linewidth,angle=90]{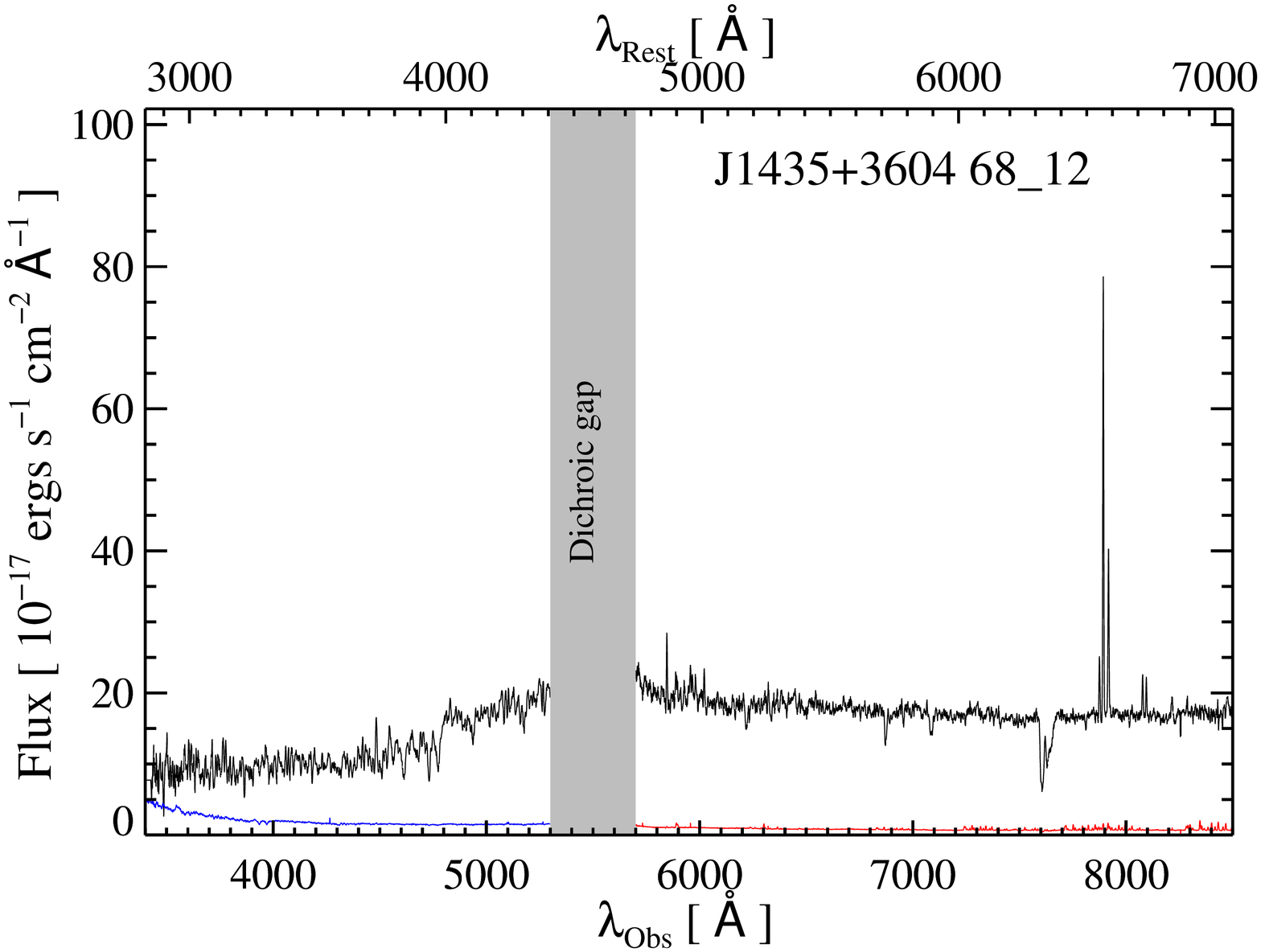} 
\end{array}$
\end{center}
\caption{The 1D reduced, flux-calibrated spectra for the galaxies included in this work. We represent the dichroic with a shaded area  near the observed wavelength 5000 \AA. Observations from \citet{Wer11}. \label{Fig:galO}}
\end{figure*}

To date, deep imaging has only been obtained for one of our absorbers. Imaging and spectroscopy of the field around J1009+0713 using the HST Wide Field Camera 3 and Keck/LRIS, respectively, was presented in \citet{Mei11}. This study revealed that none of the nearby galaxies are at the redshift of the DLA. We are limited to using the SDSS images for the other 7 sources to identify affiliated galaxies. Fortunately, spectra of several of the neighboring galaxies have been obtained with Keck/LRIS as part of a survey by \citet{Wer11}, giving us accurate redshifts for some of the galaxies in the fields of our QSO. The redshifts determined from this survey are accurate to within 30 \kms. For six of the galaxies observed, metallicities were also estimated using calibration methods of \citet{McG91} and/or the N2 index of \citet{PP04}. The properties of these galaxies are summarized in Table \ref{Tab:Gal}. The emission spectra for these galaxies are reproduced in Figure \ref{Fig:galO} from \citet{Wer11}. 

\begin{figure*}
\begin{center}$
\begin{array}{cc}
\includegraphics[width=2.3in]{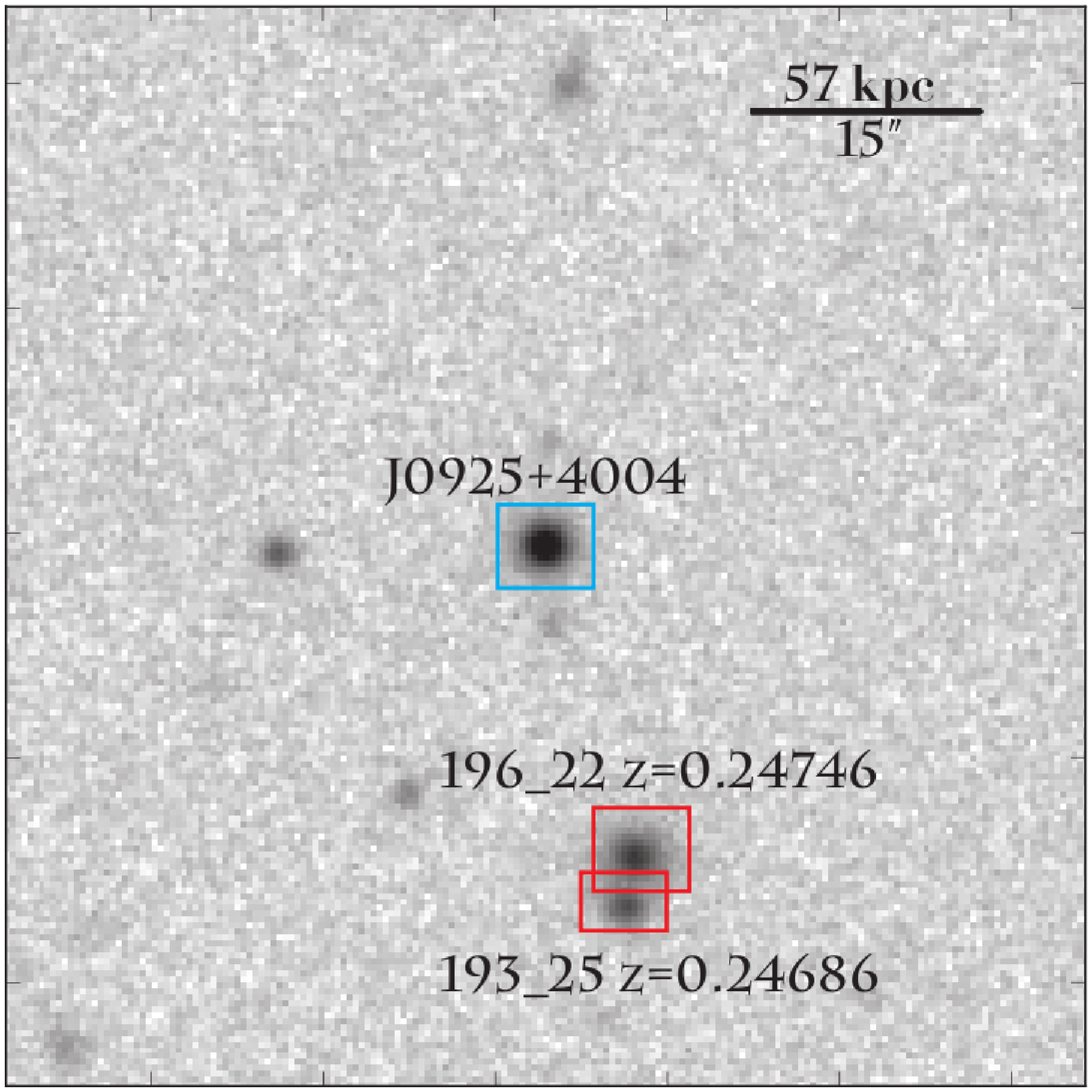} &
\includegraphics[width=2.3in]{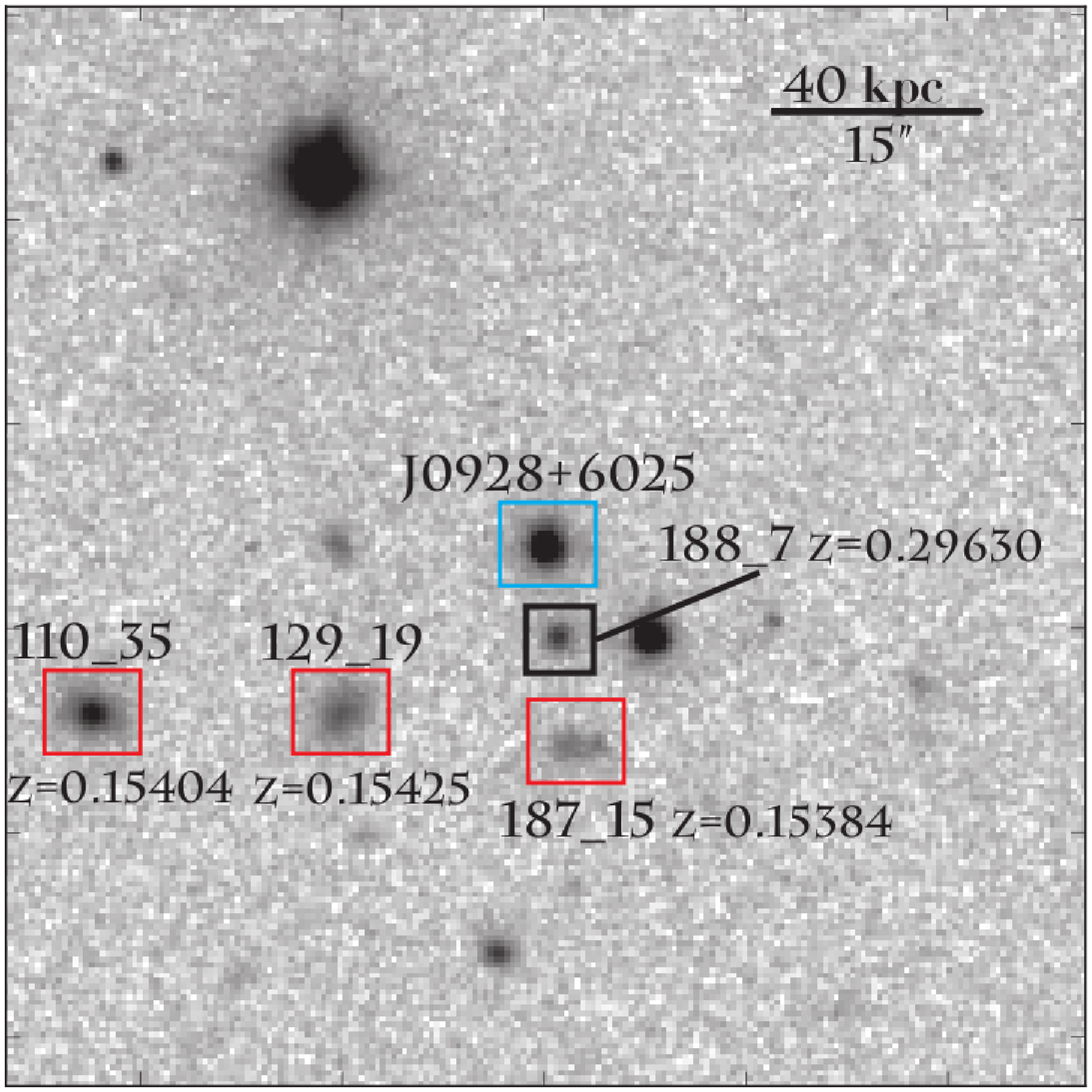} \\
\includegraphics[width=2.3in]{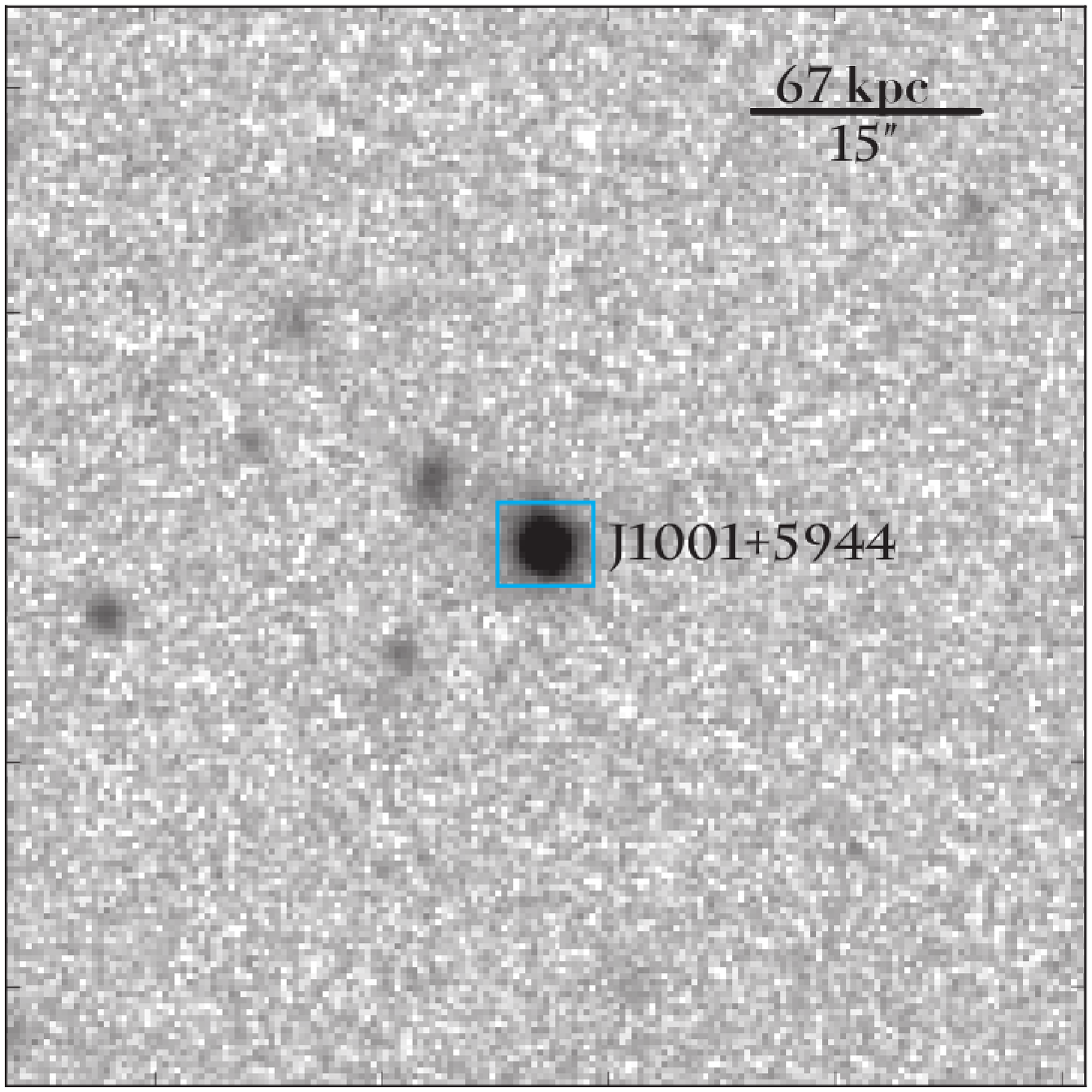} &
\includegraphics[width=2.3in]{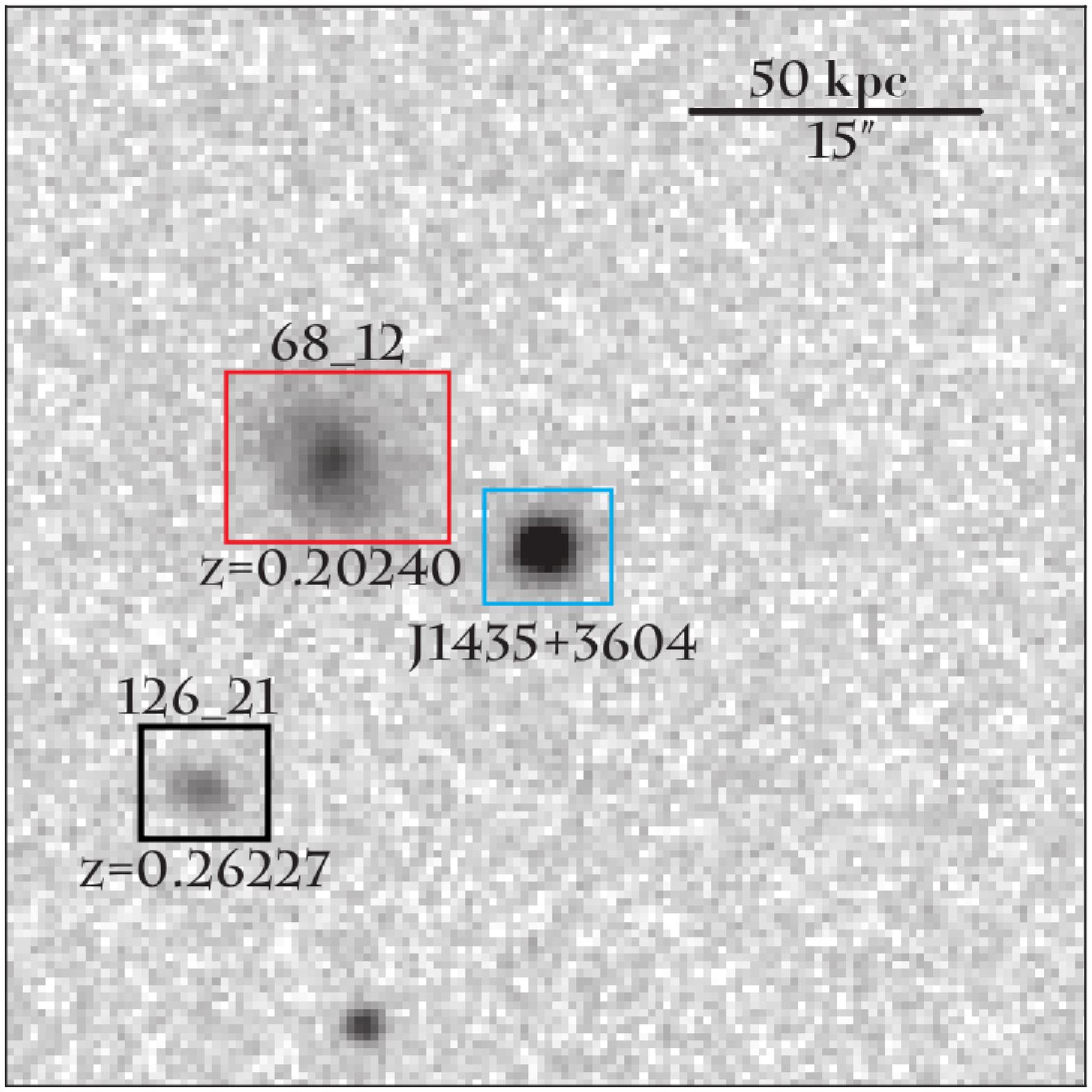} \\
\includegraphics[width=2.3in]{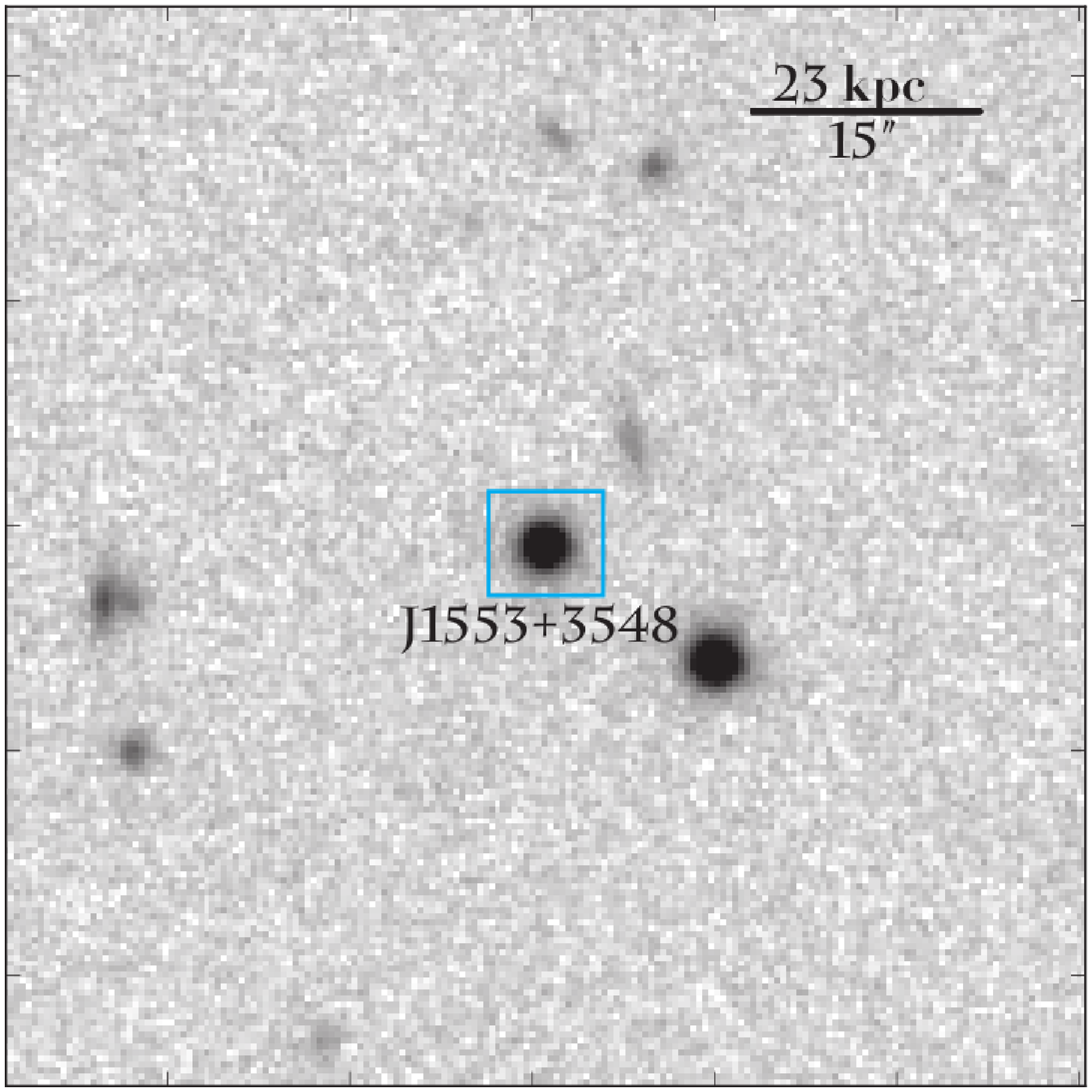} &
\includegraphics[width=2.3in]{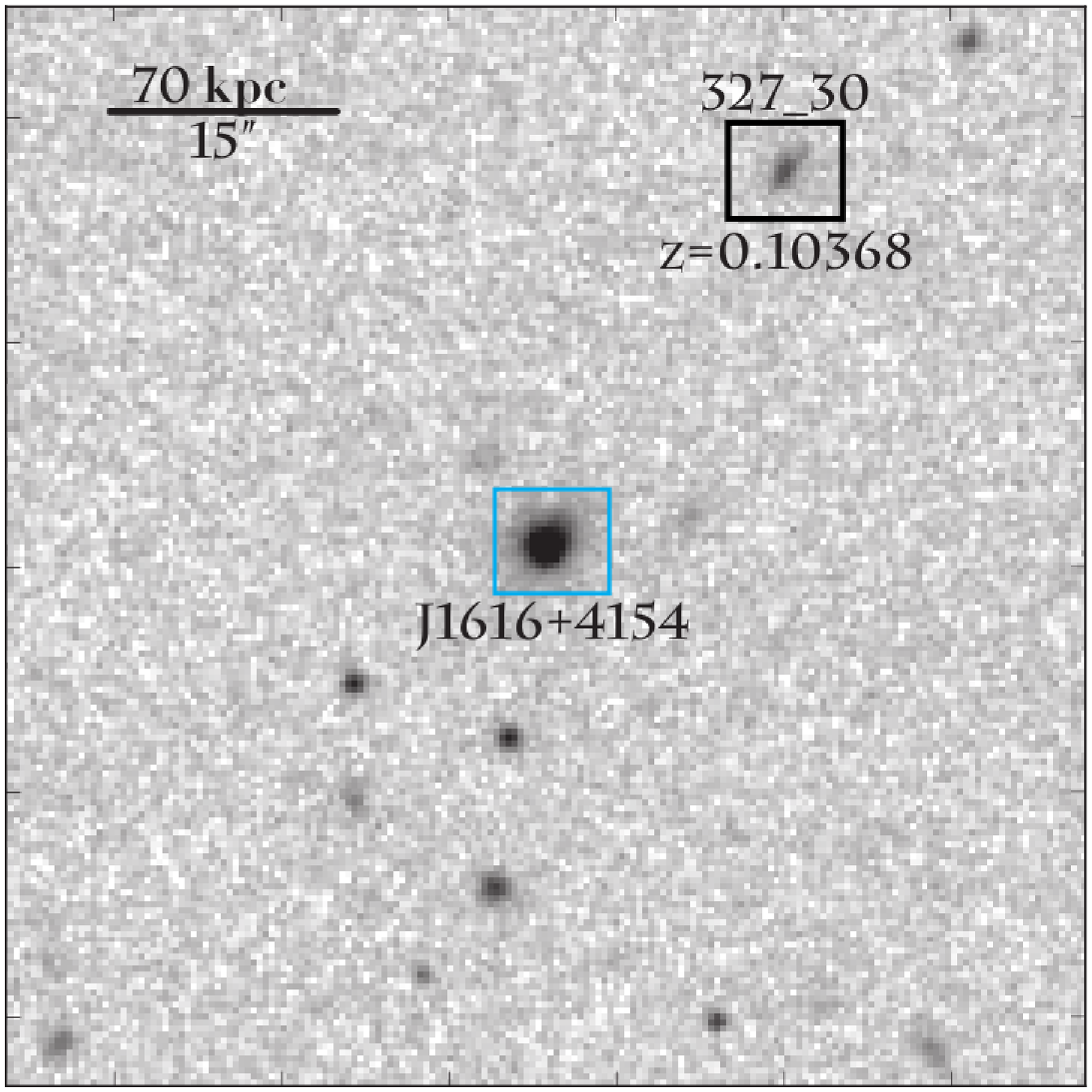} \\
\multicolumn{2}{c}{\includegraphics[width=2.3in]{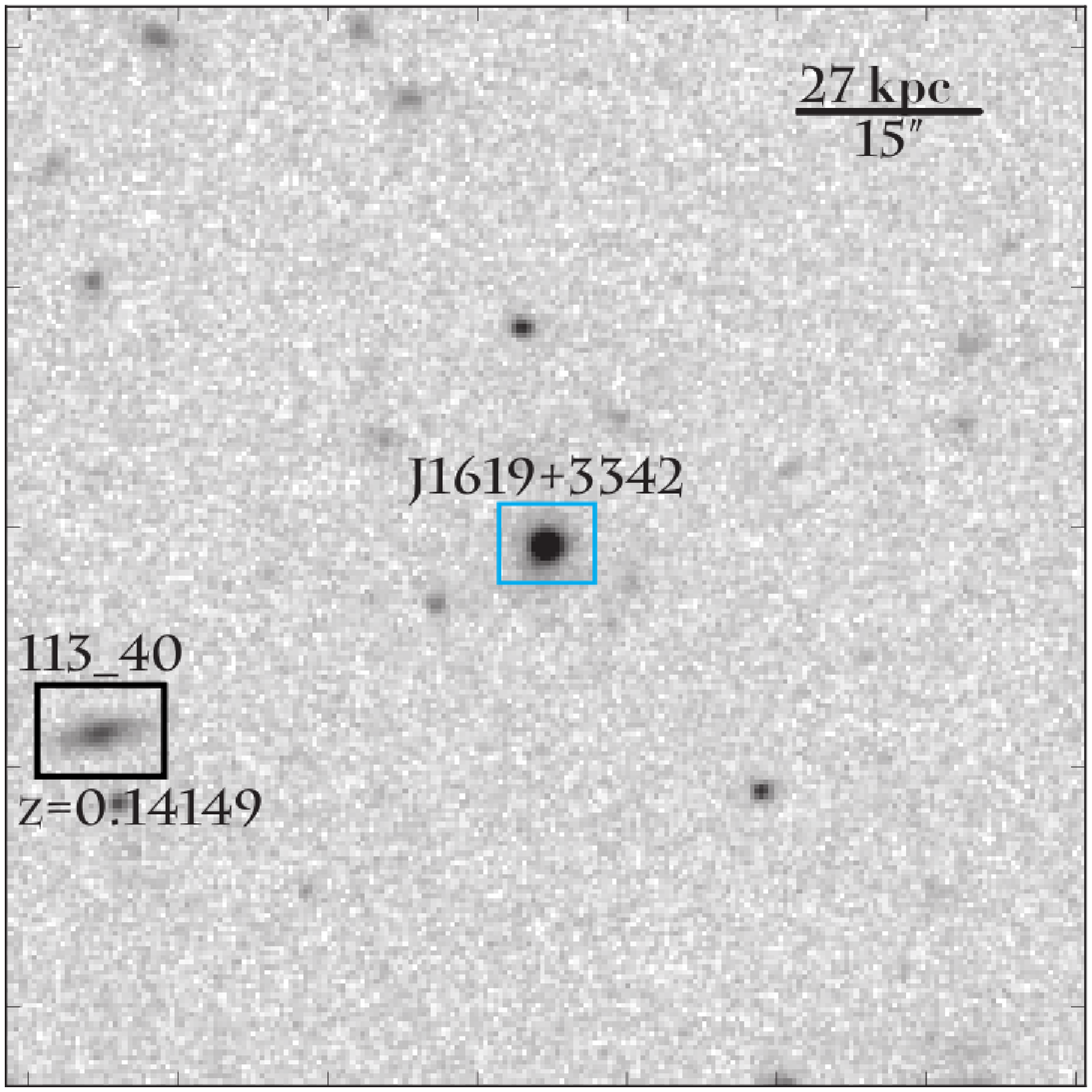}}
\end{array}$
\end{center}
\caption{SDSS images of the field around each QSO. Cyan boxes denote the QSO, red boxes denote candidate host galaxies, and black boxes are other spectroscopically observed galaxies. \label{Fig:SDSS}}
\end{figure*}

Affiliated galaxies have been found for three of the sub-DLA systems studied in this work. SDSS images of each of our absorbers, excluding J1009+0713, are shown in Figure \ref{Fig:SDSS} with candidate host galaxies indicated with a red box. Galaxies are identified using the position angle in degrees from the QSO, indicated by the first number, and the projected separation in arcseconds (impact parameter) from the QSO, indicated by the second number. The large impact parameters for all of these galaxies suggest that the sub-DLAs are likely to be in their outer halos. There are a large number of dimmer features which have yet to be observed spectroscopically that may be affiliated galaxies.

Two galaxies in the field of J0925+4004, identified as 193\_25 and 196\_22, were found to be located at a similar redshift to the sub-DLA, making them good candidates as host galaxies. Galaxies 193\_25 and 196\_22 have impact parameters of $\rho=92$ and 81 kpc, respectively, with respect to the QSO. Galaxy 196\_22 appears to be more massive than its neighbor and also more luminous, with a $L>L_*$ as determined from its $M_r$ \citep{Bla03}. The metallicity was estimated for galaxy 193\_25 using the calibration methods of \citet{McG91} to be about [O/H]$=0.12\pm0.15$, which is $\sim 0.4$ dex higher than the sub-DLA metallicity. It is well known that the metallicity in galaxies can vary from their morphological type and also from the distance from galactic center. A study by \citet{Che05} found a radial gradient of $-0.041\pm0.012$ dex kpc$^{-1}$ from the galactic center to a 30 $h^{-1}$ kpc radius for an ensemble of six galaxy-DLA pairs. However, recent work by \citet{Per11} has shown that there is a large scatter in radial gradients and also cases of positive gradients in some systems. The radial gradient necessary to explain the observed metallicity decrement from 193\_25 would be $-0.004$~dex kpc$^{-1}$, an order of magnitude lower than the rate found for the DLAs in \citet{Che05}. However, since the sub-DLA is at a considerably large impact parameter, applying a gradient seems inappropriate. Given the proximity of these galaxies, it is more likely this gas was stripped from one of these systems during an interaction.

Three galaxies in the field of J0928+6025, identified as 110\_35, 129\_19 and 187\_15, were found to be located at a similar redshift to the sub-DLA, making them good candidates as host galaxies. Galaxies 110\_35, 129\_19 and 187\_15 have impact parameters of $\rho=89$, 48 and 38 kpc, respectively, with respect to the QSO. It is worth noting that these three galaxies are part of a larger cluster centered about 750 kpc away. Galaxy 110\_35 appears to be the most massive and luminous of the three. The metallicity of galaxy 129\_19 was measured using the method of \citet{McG91} and also the N2 index of \citet{PP04} to be [O/H]$=-0.22\pm0.15$ and [O/H]$=0.07\pm0.35$, respectively. These values are below the metallicity estimated for the sub-DLA using the corrected abundance of Fe. The metallicity of galaxy 187\_15, the closest to the sub-DLA, was measured using the N2 index of \citet{PP04} to be [O/H]$=-0.38\pm0.35$, which is also lower than the sub-DLA metallicity. The complicated kinematic structure of this sub-DLA (required 13 velocity components to fit) may be the result of an interaction between these galaxies. If this is the case, the sub-DLA could be tidal debris from one of these galaxies. Observational support for such a scenario has been seen in deep HST observations by \citet{Kac10}, who observed large stellar streams and disturbed morphologies in a group of galaxies associated with a DLA, suggesting that the absorption system arises from tidal debris in the group environment.

A single galaxy in the field of J1435+3604, identified as 68\_12,  was found to be located at a similar redshift to the sub-DLA, making it a good candidate as a host galaxy.  Galaxy 68\_12 has an impact parameters of $\rho=$38 kpc with respect to the QSO. Interestingly, this galaxy shows a high star formation rate (SFR$\sim$19 M$_\odot$ yr$^{-1}$) and also appears to be very luminous, with a $L>L_*$ as determined from its $M_r$. The metallicity of galaxy 68\_12 was measured using the methods of \citet{McG91} and \citet{PP04} to be [O/H]$=0.12\pm0.15$ and [O/H]$=0.08\pm0.35$, respectively. These values are about 0.5 dex higher than the metallicity we found for our sub-DLA. The radial gradient necessary to explain this observed metallicity decrement would be $-0.013$ dex kpc$^{-1}$, or about one-third the rate found for the DLAs in \citet{Che05}. If 68\_12 is the host, the distance from galactic center may explain the metallicity decrement. However, applying a gradient may not be appropriate at such large impact parameters.

It is quite interesting that we find the potential hosts to these absorbers at such large impact parameters with the QSO. However, the fact that our candidates so far are at a relatively large impact parameter is not surprising for three reasons: (1) geometrically, it seems more likely to find an absorber intersecting the halo of a galaxy because of the relative size of the halo with respect to the disk or central region; (2) our sample was selected to be UV bright and since the QSO light will dim from extinction by the larger amounts of dust in a disk of a galaxy, it is possible those passing through a disk could be missed; (3) galaxies which are at very small impact parameters from the QSO will not show up in the SDSS images because of the brightness of the QSO (however, small-impact parameter galaxies can be identified by ``composite'' SDSS spectra showing emission from both the background QSO and the foreground galaxy, e.g., \citet{Bor10}). Deeper observations will need to be obtained to determine if there are galaxies close to the QSO sightline. It is also worth noting that log \nhI was found to be inversely correlated with $\rho$ at the 3$\sigma$ level by \citet{Rao11} and this seems consistent with only our sub-DLAs being found at large impact parameters.

From the SDSS images it seems reasonable to conclude that none of the DLAs are affiliated with $L_*$ galaxies. An $L_*$ galaxy with a diameter of 30 kpc placed at a redshift of $z_{abs}=0.1$ would be very bright, $m_i=17.1$, and have sufficient angular separation (13.4'') to be distinguishable from the large PSF of the QSO in the SDSS images. It is likely that these systems are associated with dimmer galaxies in our field that have yet to be spectroscopically observed. It is also important to perform deeper imaging of all of our fields, similar to that obtained for J1009+0713, to identify possible low-luminosity candidates at small impact parameters to the QSO. 

\begin{table*}
\begin{center}
\caption{Properties of possible host galaxies of sub-DLAs \label{Tab:Gal}}
\scalebox{1.0}{
\begin{tabular}{ccccccccccc}
\hline\hline 
Field & ID$^a$ & $z_{gal}$$^b$ & $\rho$ $^c$ & $\Delta v$$^d$ & \multicolumn{2}{c}{SFR} & \multicolumn{2}{c}{[O/H]$^e$} & Log(M$_*$) &  $M_r$ \\ 
           &         &      & (kpc)& (\kms) & Balmer       & [OII]          & M91$^f$       & PP04$^g$        &           & \\ \hline
J0925+4004 & 193\_25 & 0.2467 & 92 & -300 & 0.86$\pm$0.15  & 0.56$\pm$0.16  & 0.12$\pm$0.15 & \nodata        & 10.39 & -20.434$\pm$0.021 \\
           & 196\_22 & 0.2475 & 81 & -60  & $<$0.57        & $<$0.45        & \nodata       & \nodata        & 11.07 & -21.546$\pm$0.013 \\
J0928+6025 & 110\_35 & 0.1540 & 89 &  60  & $<$0.03        & $<$0.04        & \nodata       & \nodata        & 10.56 & -20.411$\pm$0.012 \\
           & 129\_19 & 0.1542 & 48 &  120 & 2.89$\pm$0.37  & 4.67$\pm$1.35  & -0.22$\pm$0.15 & 0.07$\pm$0.35 &  9.86 & -19.728$\pm$0.021 \\
           & 187\_15 & 0.1537 & 38 & -30  & 0.45$\pm$0.05  & 0.31$\pm$0.09  & \nodata       & -0.38$\pm$0.35 &  9.33 & -18.664$\pm$0.039 \\
J1435+3604 &  68\_12 & 0.2024 & 38 & -60  & 18.96$\pm$2.28 & 15.57$\pm$4.43 & 0.12$\pm$0.15 &  0.08$\pm$0.35 & 10.87 & -21.461$\pm$0.011 \\ \hline
\end{tabular}}
\end{center}
\textbf{Note.} Measurements taken from \citet{Wer11} \newline $^a$ Galaxy Identifier, where the first number is the position angle in degrees from the QSO and the second number is the projected separation in arcseconds (impact parameter) from the QSO. \newline $^b$ Uncertainty in redshift values is 30 \kms. \newline $^c$ Projected separation between the galaxy and QSO in kpc, calculated in the rest frame of the galaxy. \newline $^d$ Relative velocity difference of galaxy with respect to $z_{abs}$. Uncertainty is the same as redshift ($\pm$30 \kms). \newline $^e$ For reference [O/H]$_\odot=3.31$. \newline $^f$ Abundance using \citet{McG91} calibration. \newline $^g$ Abundance using the N2 index of \citet{PP04}.
\end{table*}

\section{Summary}
As part of our survey of galaxy halos with the Cosmic Origins Spectrograph, 5 sub-DLAs and 3 DLA absorption systems were serendipitously discovered. This sample nearly doubles the current sample of low-redshift QSO absorption line systems discovered to date. Accurate column densities and abundances were determined using a profile fitting procedure because of the low S/N of some of our sources and also to reduce the effects of the broad LSF of the COS spectrograph. The kinematic structure of most of the absorbers appeared to be relatively simple, requiring only a few of components over a short velocity spread. Only J0928+0625 and J1001+5944 showed complex kinematics, requiring more components spread over a larger velocity range. However, it is likely that only the Keck/HIRES spectra reflect the true component structure of the absorber and so further observations are needed to perform kinematic analysis of all of these systems.

The metallicities of our observed systems show a large scatter in metallicity from near solar to one-tenth solar. Ionization corrections were made for the sub-DLAs using the CLOUDY ionization code. However, it has been shown by \citet{Mil10} that the simplification of a single phase medium can lead to a systematic error in the result, in which the ionization correction factors could be too small. This in turn would lead to corrected abundances that may be too high. When combined with other sources from the literature, we see a trend of low-redshift DLAs being underabundant to solar metallicities. Sub-DLAs systems tend to have higher metallicities on average than DLAs at all redshifts. Unfortunately, the sampling of absorptions systems are still highly skewed toward higher redshifts. With the current sample size, only 10\% of DLAs and 23\% of sub-DLAs are at look-back times less than half the age of the Universe. For look-back times less than one-quarter of the age of the Universe, this drops to just 3\% of DLAs and 11\% of sub-DLAs. A more intuitive approach to analyze the chemical evolution of the Universe would be to divide up the observed DLAs and sub-DLAs into equal-sized age bins. The number of low-redshift absorptions systems must increase if we want to make any substantial claims as to the nature of chemical evolution. Fortunately, this task is slowly taking shape with the new COS instrument aboard HST.

For each of our low-redshift absorbers, except perhaps J1616+4154, the amount of N observed was significantly underabundant relative to $\alpha$-process elements. Analysis of the [N/$\alpha$] ratio for these systems showed that the majority lie below the transition point of secondary production levels seen in H~II regions of local spiral galaxies. If these absorbers are associated with the disks of spiral galaxies then this result seems puzzling. It is expected that such low-redshift systems would have adequate time for secondary production levels to begin. 

Observations of low redshift DLAs and sub-DLAs are proving fruitful in understanding the nature of the host galaxies. Of the 8 systems we observed, we were able to identify affiliated galaxies for 3 sub-DLAs in the SDSS images using spectra from a survey by \citet{Wer11}. The large impact parameters of these galaxies to the sub-DLAs suggest the absorbers lie in their outskirts. None of the DLAs appear associated with $L_*$ galaxies, as these would be easily observed in SDSS images. A large number of dimmer objects in our fields have yet to be measured spectroscopically and could likely yield more galaxy affiliations. More observations will be needed to properly identify the host galaxy of each of our absorption systems. Follow-up studies will ultimately improve our ability to understand the absorber-galaxy relationship and advance theories on chemical evolution and galaxy formation.

\section*{Acknowledgments}
Based on observations made with the NASA/ESA Hubble Space Telescope, obtained at the Space Telescope Science Institute, which is operated by the Association of Universities for Research in Astronomy, Inc., under NASA contract NAS 5-26555. Some of the data presented herein were obtained at the W.M. Keck Observatory, which is operated as a scientific partnership among the California Institute of Technology, the University of California and the National Aeronautics and Space Administration. The Observatory was made possible by the generous financial support of the W.M. Keck Foundation. The authors wish to extend special thanks to those of Hawaiian ancestry on whose sacred mountain we are privileged to be guests. Without their generous hospitality these observations would not have been possible. We thank M. Rafelski for sharing his table of measurements in advance of publication. The authors also wish to thank J. M. O'Meara for his helpful comments. Financial support for this research was provided by NASA grants HST-GO-11598.03-A and NNX08AJ44G. J.X.P. also acknowledges support from NSF grant (AST-0709235). 

\newpage

\appendix
\section{Column Density Measurements}
Here we breifly describe our method for using profile fits on each of the absorption systems. The velocity components and $b$ values for each system, as well as the column densities of detected species at each of these components, are shown in Table \ref{Tab:Param}.
\subsection{The Sub-DLA in SDSS J0925+4004}
The velocity centroids and $b$ values for this system were determined using O~I because we detect five lines at differing line strengths for this species. The detection of so many O~I lines is often difficult because the majority of its lines lie below 1000\AA, which is typically outside instrument coverage unless the system is significantly redshifted. The kinematics of this system appear simple with only 2 components needed to properly fit the observed profile at COS resolution. The best fit parameters from O~I were then used to fit N~I, Si~II and Fe~II, and the results can be seen in Figure \ref{Fig:J0925vel}. This system is unique among our sample as it shows numerous molecular hydrogen absorption features. These absorption lines will be analyzed in a future paper.


\begin{sidewaystable*}
\caption{Component structure and column densities for each absorber \label{Tab:Param}}
\scalebox{0.88}{
\begin{tabular}{ccccccccccccccc}
\hline\hline 
QSO & Vel & $b$ & log$N_{\rm{CII^*}}$ & log$N_{\rm{NI}}$ & log$N_{\rm{NII}}$ & log$N_{\rm{OI}}$ & log$N_{\rm{MgI}}$ &log$N_{\rm{SiII}}$ & log$N_{\rm{PII}}$ & log$N_{\rm{SII}}$ & log$N_{\rm{CaII}}$ & log$N_{\rm{TiII}}$ &  log$N_{\rm{FeII}}$ & log$N_{\rm{FeIII}}$ \\ 
 & (\kms) & (\kms) & (cm$^{-2}$) & (cm$^{-2}$) & (cm$^{-2}$) & (cm$^{-2}$) & (cm$^{-2}$) & (cm$^{-2}$) & (cm$^{-2}$) & (cm$^{-2}$) & (cm$^{-2}$) & (cm$^{-2}$) & (cm$^{-2}$) & (cm$^{-2}$)\\ \hline
J0925+4004 & -34$\pm$4 & 24$\pm$4 & \nodata & 14.14$\pm$0.06 & \nodata &  15.36$\pm$0.06 & \nodata & 14.32$\pm$0.07 & \nodata & \nodata & \nodata & \nodata & 13.16$\pm$0.76 & \nodata \\
 & 41$\pm$2 & 20$\pm$3 & \nodata & 14.63$\pm$0.05 & \nodata &  15.82$\pm$0.11 & \nodata & 14.32$\pm$0.09 & \nodata & \nodata & \nodata & \nodata & 14.04$\pm$0.11 & \nodata \\ \hline
J0928+6025 & -39.5$\pm$0.6 & 6.3$\pm$0.9 & \nodata & \nodata & \nodata & \nodata & 12.00$\pm$0.05 & \nodata & \nodata & \nodata & \nodata & \nodata & \nodata & \nodata \\
 & -29.6$\pm$0.5 & 0.5$\pm$0.5 & \nodata & \nodata & \nodata & \nodata & 10.96$\pm$2.42 & \nodata & \nodata & \nodata & \nodata & \nodata & \nodata & \nodata \\
 & -24.4$\pm$1.1 & 2.1$\pm$1.7 & \nodata & \nodata & \nodata & \nodata & 11.53$\pm$0.23 & \nodata & \nodata & \nodata & \nodata & \nodata & \nodata & \nodata \\
 & -19.0$\pm$2.0 & 2.0$\pm$1.9 & \nodata & \nodata & \nodata & \nodata & 11.32$\pm$0.98 & \nodata & \nodata & \nodata & \nodata & \nodata & \nodata & \nodata \\
 & -13.5$\pm$3.6 & 4.0$\pm$3.1 & \nodata & \nodata & \nodata & \nodata & 11.56$\pm$0.49 & \nodata & \nodata & \nodata & \nodata & \nodata & \nodata & \nodata \\
 & 3.5$\pm$1.0 & 8.7$\pm$2.1 & \nodata & \nodata & \nodata & \nodata & 12.15$\pm$0.07 & \nodata & \nodata & \nodata & \nodata & \nodata & \nodata & \nodata \\
 & 14.5$\pm$1.0 & 2.1$\pm$1.6 & \nodata & \nodata & \nodata & \nodata & 11.25$\pm$0.37 & \nodata & \nodata & \nodata & \nodata & \nodata & \nodata& \nodata  \\
 & 28.1$\pm$9.4 & 8.5$\pm$7.2 & \nodata & \nodata & \nodata & \nodata & 11.59$\pm$0.66 & \nodata & \nodata & \nodata & \nodata & \nodata & \nodata & \nodata \\
 & 36.5$\pm$0.5 & 3.8$\pm$1.1 & \nodata & \nodata & \nodata & \nodata & 11.98$\pm$0.23 & \nodata & \nodata & \nodata & \nodata & \nodata & \nodata & \nodata \\ \hline
J1001+5944 & -38$\pm$1 & 20$\pm$2 & \nodata & 13.44$\pm$0.09 & \nodata &  15.44$\pm$0.02 & \nodata & 14.50$\pm$0.04 & 12.75$\pm$0.07 & \nodata & \nodata & \nodata & 14.05$\pm$0.05 & 13.88$\pm$0.08 \\
 & 20$\pm$2 & 18$\pm$4 & \nodata & 13.06$\pm$0.19 & \nodata &  15.15$\pm$0.03 & \nodata & 14.26$\pm$0.04 & \nodata & \nodata & \nodata & \nodata & 13.88$\pm$0.06 & 13.62$\pm$0.14 \\
 & 63$\pm$18 & 13$\pm$13 & \nodata & \nodata & \nodata &  13.86$\pm$0.14 & \nodata & 13.08$\pm$0.07 & 11.93$\pm$0.33 & \nodata & \nodata & \nodata & 13.07$\pm$0.33 & 13.17$\pm$0.34 \\
 & 92$\pm$6 & 22$\pm$9 & \nodata & 12.50$\pm$0.71 & \nodata &  14.01$\pm$0.11 & \nodata & 13.43$\pm$0.04 & \nodata & \nodata & \nodata & \nodata & \nodata & \nodata  \\ \hline
J1435+3604 & -31$\pm$2 & 9$\pm$4 & \nodata & 13.58$\pm$0.08 & \nodata &  \nodata & \nodata & \nodata & \nodata & 14.00$\pm$0.23 & \nodata & \nodata & 13.54$\pm$0.20 & \nodata \\
 & -4$\pm$1 & 5$\pm$1 & \nodata & 14.56$\pm$0.16 & \nodata &  \nodata & \nodata & \nodata & \nodata & 14.47$\pm$0.15 & \nodata & \nodata & 14.36$\pm$0.12 & \nodata \\ \hline
J1553+3548 & -29$\pm$1 & 29$\pm$4 & \nodata & \nodata & 14.16$\pm$0.07 & \nodata & \nodata & 14.22$\pm$0.05 & \nodata &  \nodata & \nodata & \nodata & 14.01$\pm$0.07 & \nodata \\ \hline
J1616+4154 & -106$\pm$20 & 20$\pm$15 & \nodata & \nodata & \nodata & \nodata & \nodata & \nodata & \nodata & \nodata & \nodata & \nodata & 13.79$\pm$0.61 & \nodata \\
 & -59$\pm$10 & 28$\pm$16 & 13.43$\pm$0.17 & \nodata & \nodata & \nodata & \nodata & \nodata & 12.39$\pm$0.98 & 14.57$\pm$0.47 & \nodata & \nodata & 14.32$\pm$0.23 & 14.24$\pm$0.06 \\
 & 10$\pm$2 & 24$\pm$2 & 13.79$\pm$0.09 & \nodata & \nodata & \nodata & \nodata & \nodata & 13.42$\pm$0.10 & 15.29$\pm$0.12 & \nodata & \nodata & 14.89$\pm$0.03 & 14.18$\pm$0.06 \\ \hline
J1619+3342 & -29.8$\pm$2.5 & 4.4$\pm$2.1 & \nodata & \nodata & \nodata & \nodata & 11.89$\pm$0.26 & \nodata & \nodata & \nodata & 11.55$\pm$0.04 & 11.11$\pm$0.19 & \nodata & 13.45$\pm$0.37\\
 & -21.0$\pm$1.4 & 4.1$\pm$1.2 & \nodata & 14.63$\pm$0.12 & \nodata & \nodata & 12.24$\pm$0.18 & \nodata & 13.17$\pm$0.19 & 15.08$\pm$0.09 & 12.36$\pm$0.01 & 11.83$\pm$0.04 & 14.25$\pm$0.16 & 13.78$\pm$0.28 \\ \hline
\end{tabular}}
\newline \newline \textbf{Note.} This table is just component-by-component column densities from the profile fit and does not include AOD column density estimates or limits.
\end{sidewaystable*}

\subsection{The Sub-DLA in SDSS J0928+6025}
Optical spectra of J0928+6025 were obtained with the Keck/HIRES spectrograph. The absorption lines detected by Keck for this system are shown in Figure \ref{Fig:J0928keckvel}. The velocity centroids and $b$ values for this system were determined using Mg~I. The kinematic structure of this system is quite complex and required 9 components to properly fit the observed profile. An additional 4 components were necessary to fit weaker outer components appearing in the much stronger Mg~II lines. The Ca~II line was independently fit, as it shows slightly different kinematics than Mg~I, and the result is presented in Table \ref{Tab:J0928Cavel}. Since the resolution of the COS instrument is not high enough to distinguish all the components found in Mg~I, the attempted COS fits utilized only the components at $v=-39.52$,$-18.95$, $3.5$, $36.51$ \kms, as these appear to be the strongest features in the Keck data. Unfortunately, the lines observed in the COS data were too strongly saturated or were too noisy for reasonable fits to be made of the observed profile. This can be seen in Figure \ref{Fig:J0928vel}. Therefore, apparent column densities are adopted for each line in the COS data. For this case, lines were determined to be unsaturated if the apparent column densities of two or more lines with different line strengths were consistent.


\begin{table}
\caption{Parameter fit to Ca~II in the $z_{abs}$=0.1538 SDSS J0928+6025 sub-DLA \label{Tab:J0928Cavel}}
\begin{center}
\begin{tabular}{cccc}
\hline\hline 
Vel & $b$ & log$N_{\rm{CaII}}$ \\  
(\kms) & (\kms) & (cm$^{-2}$) \\\hline
-39.8$\pm$0.3 & 4.8$\pm$0.5 & 11.98$\pm$0.03 \\
-29.4$\pm$0.9 & 3.2$\pm$1.6 & 11.38$\pm$0.20 \\
-21.5$\pm$0.5 & 3.5$\pm$1.0 & 11.80$\pm$0.11 \\
-12.5$\pm$0.7 & 4.5$\pm$1.3 & 11.75$\pm$0.11 \\
0.2$\pm$2.2 & 4.7$\pm$1.9 & 11.90$\pm$0.32 \\
8.4$\pm$2.6 & 5.2$\pm$2.6 & 11.92$\pm$0.32 \\
18.2$\pm$0.8 & 2.4$\pm$1.3 & 11.26$\pm$0.24 \\
31.7$\pm$2.1 & 8.9$\pm$1.7 & 11.89$\pm$0.12 \\
36.4$\pm$0.2 & 3.3$\pm$0.4 & 12.18$\pm$0.06 \\\hline
\end{tabular}
\end{center}
\end{table}

\subsection{The Sub-DLA in SDSS J1001+5944}
The velocity centroids and $b$ values for this system were determined using O~I because of the large number of observed lines at differing line strengths. The kinematics of this system appear complex with 4 components over a large velocity spread being needed to properly fit the observed profile at COS resolution. The O~I parameters were then used to fit N~I, Si~II, PII, and Fe~II, and the results can be seen in Figure \ref{Fig:J1001vel}. This sub-DLA shows a Lyman limit break in the continuum at a wavelength of 1195\ang in the COS rest frame.


\subsection{The Sub-DLA in SDSS J1435+3604}
The velocity centroids and $b$ values for this system were determined using N~I because there are 6 observed lines and most of the O~I lines appear saturated or blended. The kinematics of this system appear simple with only 2 components needed to properly fit the observed profile at COS resolution. The velocity parameters from the N~I fit were then used to fit Si~II, S~II, and Fe~II, and the results can be seen in Figure \ref{Fig:J1435vel}.


\subsection{The Sub-DLA in SDSS J1553+3548}
The velocity centroids and $b$ values for this system were determined using Si~II since it was the most extensively covered line. The kinematics of this system appear very simple with only a single component needed to properly fit the observed profile at COS resolution. The Si~II parameters were then used to fit N~II and Fe~II, and the results can be seen in Figure \ref{Fig:J1553vel}.


\subsection{The DLA in SDSS J1616+4154}
The velocity centroids and $b$ values for this system were determined using Fe~II. The kinematics of this system appear simple with only 3 components needed to properly fit the observed profile at COS resolution. The parameters from the Fe~II fit were then used to fit C~II*, P~II, S~II, and Fe~II and the results can be seen in Figure \ref{Fig:J1616vel}.


\subsection{The DLA in SDSS J1619+3342}
Optical spectra of J1619+3342 were obtained with the Keck/HIRES spectrograph. The absorption lines detected by Keck for this system are shown in Figure \ref{Fig:J1619keckvel}. The Mg~II lines lie at the edge of the detector and so have very low S/N. The velocity centroids and $b$ values for this system were determined using Mg~I. The kinematic structure of this system is simple and required only 2 components to properly fit the observed profile. For very strong transitions, such as Mg~II, C~II, and Si~III, an additional component at $v=-143.0$ \kms\ seems to appear. However, this additional component was not used in the fits of the weaker lines. The Mg~I fit parameters were then used to fit N~I, Si~II, P~II, S~II, Ca~II, Fe~II, and Fe~III and the results can be seen in Figure \ref{Fig:J1619vel}. Since the resolution of the COS instrument is lower than that of Keck, some of the COS fits utilize only the component at $v=-21.0$ \kms, since it appears to contain most of the column density.


\end{document}